\documentclass[%
amsmath,amssymb,
preprint,
11pt,
preprint,
aps,
fleqn
]{revtex4-1}

\usepackage{bm}
\usepackage[dvipdfmx]{graphicx}
\usepackage[usenames]{color}
\usepackage[normalem]{ulem}

\begin{document}

\title{Phase reduction theory for hybrid nonlinear oscillators}

\author{Sho Shirasaka}
\email{Corresponding author: shirasaka.s.aa@m.titech.ac.jp}
\affiliation{
Graduate School of Information Science and Engineering,
Tokyo Institute of Technology, O-okayama 2-12-1, Meguro, Tokyo 152-8552, Japan
}
\author{Wataru Kurebayashi}
\affiliation{
Faculty of Software and Information Technology, Aomori University, Kobata 2-3-1, Aomori, Aomori 030-0943, Japan
}
\author{Hiroya Nakao}
\affiliation{
School of Engineering,
Tokyo Institute of Technology, O-okayama 2-12-1, Meguro, Tokyo 152-8552, Japan
}

\date{\today}

\begin{abstract}
Hybrid dynamical systems characterized by discrete switching of smooth dynamics have been used to model various rhythmic phenomena. However, the phase reduction theory, a fundamental framework for analyzing the synchronization of limit-cycle oscillations in rhythmic systems, has mostly been restricted to smooth dynamical systems.
Here we develop a general phase reduction theory for weakly perturbed limit cycles in hybrid dynamical systems that facilitates analysis, control, and optimization of nonlinear oscillators whose smooth models are unavailable or intractable. 
On the basis of the generalized theory, we analyze injection locking of hybrid limit-cycle oscillators by periodic forcing and reveal their characteristic synchronization properties, such as ultrafast and robust entrainment to the periodic forcing and logarithmic scaling at the synchronization transition. We also illustrate the theory by analyzing the synchronization dynamics of a simple physical model of biped locomotion.

\end{abstract}

\pacs{Valid PACS appear here}

\maketitle

\section{Introduction}
Hybrid dynamical systems have been used to describe physical processes that exhibit sudden qualitative changes or abrupt jumps during otherwise continuous evolution.
Some examples are the collision of particles, refraction and reflection of waves, spiking of neurons, switching of gene expression, limb-substrate impacts in legged robots and animals, human-structure interaction, switching of elements in electric circuits, and breakdown of nodes or links in networked systems~\cite{Aranson2006patterns,Wolf99geometry,Aihara2010theory,Holmes2006dynamics,Macdonald09lateral,Andronov1987theory,helbing2003modelling+Hespanha2001hybrid+Susuki09hybrid}.
Because such discontinuous events are found in many areas of science and engineering~\cite{Makarenkov2012dynamics,Bernardo08Piecewise}, it is important to
develop theoretical frameworks to analyze hybrid dynamical systems.

Many hybrid dynamical systems exhibit stable rhythmic activities, for example, periodic spiking of neurons,
rhythmic locomotion of robots, oscillations in power electric circuits, and business cycles in economic models~\cite{coombes2012nonsmooth,McGeer90+Zimmermann07,Banerjee01,jenkins2013self}, which are typically modeled as nonlinear limit-cycle oscillations. 
Synchronization of such rhythmic activities may play important functional roles in biological and engineered systems,
e.g., in locomotor rhythms, vibro-impact energy harvesters, and wireless sensor networks~\cite{Ijspeert98central+Strogatz05,Dunnmon11power+Moss10low,tanaka2009self}.

One of the fundamental mathematical frameworks for analyzing limit-cycle oscillations is the phase reduction theory~\cite{winfree2001geometry,kuramoto2003chemical,hoppensteadt1997weakly},
which gives approximate reduced description of the dynamics of a weakly perturbed limit-cycle oscillator using a simple one-dimensional phase equation.
The phase reduction theory is well established for stable limit-cycle oscillations of smooth dynamical systems and has successfully been applied to the analysis of rhythmic spatiotemporal dynamics in chemical and biological systems~\cite{winfree2001geometry,kuramoto2003chemical}. 
Methods for optimizing and controlling synchronization of limit-cycle oscillators have also been developed on the basis of the phase reduction theory~\cite{Dorfler14synchronization+Harada10optimal+Zlotnik13optimal+Zlotnik16phaseselective}.  

In phase reduction theory for smooth dynamical systems, a weakly perturbed limit-cycle oscillator described by
$\dot{{\bm X}}(t)={\bm F}({\bm X}(t))+\epsilon {\bm p}({\bm X}(t),t)$
 is considered,
where ${\bm X}(t)\in\mathbb{R}^{N}$ is the oscillator state, 
${\bm F}({\bm X}):\mathbb{R}^{N}\to \mathbb{R}^{N}$ is a continuously differentiable vector field representing the dynamics of the oscillator, ${\bm p}({\bm X},t): \mathbb{R}^{N}\times \mathbb{R}\to \mathbb{R}^{N}$ denotes external perturbation applied to the oscillator, and $\epsilon \in \mathbb{R}$ is a small parameter representing the intensity of the perturbation.
A system without perturbation $(\epsilon=0)$ is assumed to possess a stable limit-cycle solution $\chi:{\bm X}_0(t) = {\bm X}_0(t+T)$ of period $T\in\mathbb{R}$, and a phase $\theta$ of the oscillator state is introduced, which increases with a constant frequency and takes the same value on the {\it isochron}~\cite{winfree2001geometry,hale69,guckenheimer75}, i.e., a codimension-one manifold of the oscillator states that share the same asymptotic behavior.

When the perturbation is sufficiently small,
phase reduction theory enables us to systematically approximate the original multidimensional system using a simple one-dimensional reduced phase equation of the form
$\dot{\theta}(t)=1+\epsilon {\bm Z}(\theta) \cdot {\bm p}(\theta,t),$
where $\theta(t) = \Theta({\bm X}(t))$ is the oscillator phase and 
$\Theta({\bm X}):\mathbb{R}^N \to [0,T)$
gives the phase of the oscillator state ${\bm X}$. 
The range $[0, T)$ of the phase is identified with a one-dimensional torus $\mathbb{T}^1$.
The function ${\bm Z}(\theta): \mathbb{T}^1 \to \mathbb{R}^n$, which is the gradient of the isochron and is called the {\it phase sensitivity function} in this study, quantifies the linear response of the oscillator phase to perturbations given at phase $\theta$ on $\chi$.
It is known that ${\bm Z}(\theta)$ can be obtained as a $T$-periodic solution to the adjoint linear problem of the system, $\dot{{\bm Z}}(t)=-\left( D{\bm F} ({\bm X}_0 (t))\right)^\mathrm{\dag} \cdot {\bm Z}(t)$ with a normalization condition ${\bm Z}(0)\cdot {\bm F}({\bm X}_0(0))= 1$, where $D{\bm F}$ denotes the Jacobi matrix of ${\bm F}$ and $\dag$ its transpose~\cite{malkin1949methods,hoppensteadt1997weakly,ermentrout2010mathematical}.

Recently, the phase reduction theory has been extended to nonconventional cases such as stochastic~\cite{Yoshimura08+Teramae09+Goldobin10}, 
delay-induced~\cite{Novicenko12+Kotani12}, collective~\cite{kawamurai2008collective+kori2009collective},
spatially extended~\cite{kawamura2013+nakao2014},
and strongly modulated~\cite{kurebayashi2013phase} oscillations.
Similar reduction methods that rely on sets of initial conditions characterized by the same long-term behavior have also been developed for heteroclinic orbits~\cite{shaw2012phase}, limit tori~\cite{demir2010+kawamura2015+mauroy2012} and stable fixed points~\cite{Ichinose98+Rabinovich99+Mauroy13}.
However, application of the phase reduction theory to oscillatory hybrid dynamical systems has so far been limited to low-dimensional systems, or to a specific class of systems whose phase sensitivity function is obtained from adiabatic approximation~\cite{coombes2012nonsmooth,izhikevich2000phase,park2013infinitesimal}. To the best of our knowledge, no systematic phase reduction theory for oscillators of high-dimensional ($N \ge 3$) systems with discontinuity in $\chi$ has been developed. 
This is mainly because the nonsmoothness of the vector fields at the jumps prevents straightforward utilization of the adjoint equation.

In this study, we develop a systematic phase reduction theory for a general class of autonomous limit-cycle oscillators in hybrid dynamical systems. 
This paper is organized as follows: in Sec. II, limit-cycle oscillations in hybrid dynamical systems are introduced. In Sec. III, the phase reduction theory for hybrid limit cycles is developed. In Sec. IV, the theory is illustrated by analyzing synchronization dynamics of two examples of hybrid limit-cycle oscillators, that is, an analytically tractable Stuart-Landau-type oscillator and a physical model of biped locomotion. Section V summarizes the results, and Appendices A-I provide mathematical details of the main results presented in Sec. II-IV. 

\section{HYBRID LIMIT CYCLES}

The state of a hybrid dynamical system that we consider in this study is represented by a pair ${\bm s} = (I, {\bm X})$ of the discrete state $I \in \{1,2, \cdots, m \}=\mathcal{M}$ for some $m \in \mathbb{N}$ ($m = +\infty$ is allowed) and the continuous state ${\bm X} \in \mathbb{R}^{N}$.
We denote the set of pairs $(i,j)$ as $\mathcal{G} \subset \mathcal{M}\times \mathcal{M}$, which is a collection of all possible transitions from discrete state $i$ to $j$.
As in Ref.~\cite{khan2011sensitivity}, we describe a hybrid dynamical system by the following hybrid automaton:
\begin{align}
	&\dot{{\bm X}}(t) = {\bm F}(I(t),{\bm X}(t)),
	\mbox{\ \ if\ \ } \ I(t) = i \ \mbox{and}\ {\bm X}(t) \notin {\bm \Pi}_{ij} \ \mbox{for any}\ j, \label{eq. hybDS}
\end{align}
\begin{align}
	&{\bm X}(t+0) = {\bm \Phi}((i,j),{\bm X}(t)),\ I(t+0) = j, 
	\mbox{\ \ if\ \ } I(t) = i \mbox{ and } {\bm X}(t) \in {\bm \Pi}_{ij} \ \mbox{for some}\ j, \label{eq. resetF}
\end{align}
\begin{align}
	{\bm \Pi}_{ij} = \begin{cases}
		\{ {\bm X}\; | \; L((i,j),{\bm X})=0 \}, \mbox{\ \ if\ \ } (i,j)\in \mathcal{G}. \\
		\mbox{empty set, \ \ otherwise}.
	\end{cases}
		\label{eq. resetplane}
\end{align}
Here, Eq.~(\ref{eq. hybDS}) describes the smooth dynamics of the continuous state ${\bm X}(t)$ when the discrete state is $I(t) = i$,
Eq.~(\ref{eq. resetF}) the jump of ${\bm X}(t)$ when the discrete state switches from $i$ to $j$, and
Eq.~(\ref{eq. resetplane}) represents a plane in the space of continuous state on which the switching from $i$ to $j$ takes place.
In Eq.~(\ref{eq. hybDS}), ${\bm F}(I, {\bm X}): \mathcal{M}\times \mathbb{R}^{N} \to \mathbb{R}^{N}$ is the vector field of the system.
In Eq.~(\ref{eq. resetF}), ``$t+0$'' indicates the moment just after the switching of the discrete state at $t$, the transition function ${\bm \Phi}((i,j),{\bm X}):\mathcal{G}\times \mathbb{R}^{N}\to \mathbb{R}^{N}$ gives the new continuous state after the switching of the discrete state from $i$ to $j$, and ${\bm \Pi}_{ij}$ is an $(N-1)$ dimensional zero-level surface of the function $L((i,j),{\bm X}):\mathcal{G} \times \mathbb{R}^{N} \to \mathbb{R}$ on which the switching takes place. 
It is assumed that the functions ${\bm F}(I,{\bm X})$, ${\bm \Phi}( (i,j),{\bm X})$, and $L\left( (i,j),{\bm X} \right)$ are continuously differentiable with respect to ${\bm X} \in \mathbb{R}^{N}$ and do not depend explicitly on time.

Suppose there exists a periodic solution $\chi: {\bm s}_0(t) = (I_{0}(t), {\bm X}_0(t))$ of period $T$ of Eqs.~(\ref{eq. hybDS})-(\ref{eq. resetplane}). 
As in Ref.~\cite{Akhmet2005on+akhmet2010principles}, we make several assumptions on the system (see Appendix~\ref{sec. assumptions} for details) so the continuous part of the solution ${\bm X}_0(t)$ is piecewise continuously differentiable with respect to the initial continuous state ${\bm X}_0(0)$ on $\chi$ and linear stability analysis of the solution can be performed.
Let ${\bm s}^*=(I(0),{\bm X}(0))$ be an initial condition of Eqs.~(\ref{eq. hybDS})-(\ref{eq. resetplane}) and $t=\tau_k({\bm s}^*)$, $k\in \mathbb{N}$ be the moments of switching of the discrete state, where $0\le \tau_1({\bm s}^*)< \tau_2({\bm s}^*)< \cdots < \tau_{k}({\bm s}^*) < \cdots < +\infty$. For convenience of notation, we also define $\tau_0({\bm s}^*)=0$.

To simplify the expression of the periodic orbit $\chi$, we hereafter use the following notation. By distinguishing the discrete states visited more than once in one period, and by renumbering the state indices, we introduce a set of discrete states $\mathcal{M}_0 = \{1,2, \cdots , m_0 \}$ where $m_0 < + \infty$, such that the discrete state $I_{0}(t)$ is switched in numerical order as $1 \to 2 \to \cdots \to m_{0} \to m_{0}+1 = 1$ at $t=\tau_1({\bm s}^*), \tau_2({\bm s}^*), \cdots , \tau_{m_0}({\bm s}^*)$ (see Fig.~\ref{fig. schematics}). 
Here, $m_0$ is finite because the period $T$ is finite and the assumption (C2) in Appendix~\ref{sec. assumptions} assures that the system stays in each discrete state for some nonzero duration. 
We also introduce the following simplified notations for the discrete state transitions on the periodic orbit $\chi$: 
\begin{align}
	&L_k({\bm X}_0(t)) = L( (k,k+1),{\bm X}_0(t)),
	\cr
	&{\bm \Phi}_k({\bm X}_0(t)) = {\bm \Phi}( (k,k+1 ),{\bm X}_0(t)).
\end{align}
We call the periodic solution $\chi$ a hybrid limit cycle if it is linearly stable (see Appendix~\ref{sec. linearstability} for the linear stability analysis of the periodic solution).

\begin{figure}
	\includegraphics[width=\textwidth]{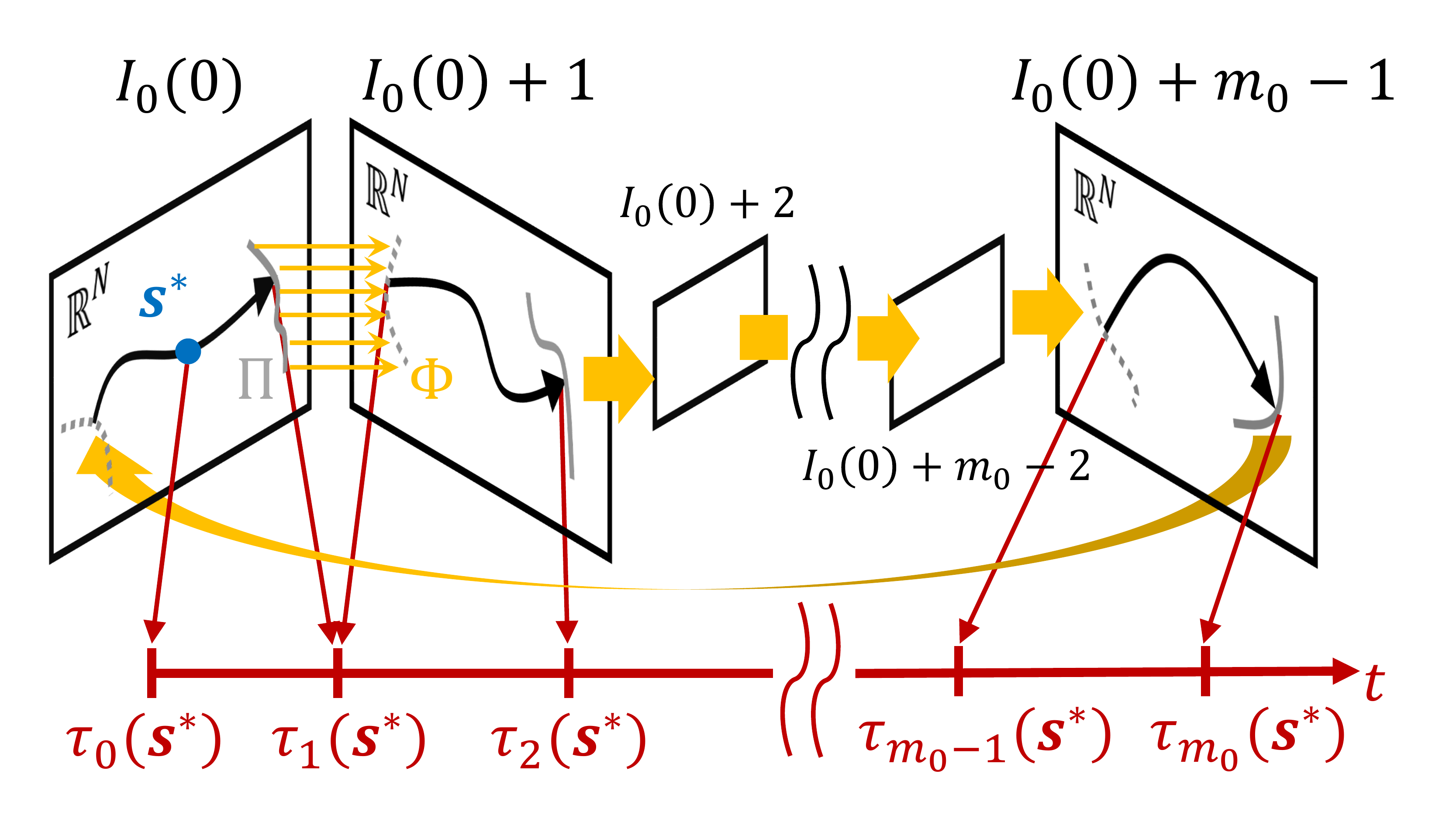}
	\caption{Schematic representation of the dynamics of the hybrid limit cycles.}
	\label{fig. schematics}
\end{figure}

\section{Phase reduction}

The aim of phase reduction is to describe the dynamics of the system state around the hybrid limit cycle $\chi$ by using a scalar phase $\theta$.
We first introduce the phase function $\Theta : {\mathcal M}_{0} \times {\mathbb R}^{N} \to {\mathbb T}^1 ( = [0,T))$ on $\chi$, which gives the phase value of the state ${\bm s}_0 = (I_{0}, {\bm X}_{0})$ on $\chi$ as
\begin{align}
	&\Theta({\bm s}_0(t+nT)) 
	=
	\Theta(I_{0}(t+ nT), {\bm X}_0(t+ nT)) = t \ (\mbox{mod } T),
	\label{eq. phaseonchi}
\end{align}
where $n\in \mathbb{Z}_{\ge 0}$ is an arbitrary positive integer. 
Namely, we identify the time $t$ (mod $T$) as the oscillator phase $\theta$, which increases with a constant frequency $1$ on $\chi$, i.e., 
\begin{align}
\dot{\theta}(t) = \dot{\Theta}(I_{0}(t), {\bm X}_{0}(t)) \equiv 1.
\end{align}
In the following, we will denote a system state with phase $\theta$ on $\chi$ also as ${\bm s}_0 (\theta) = (I_{0}(\theta), {\bm X}_{0}(\theta))$ as a function of $\theta$.

Phase $\theta$ can also be introduced in a neighborhood $U$ containing $\chi$ within its basin of attraction
by introducing an equivalence relation to initial conditions in $U$ whose asymptotic behaviors are the same.
Namely, we introduce the {\it isochron} of $\chi$ by assigning the same phase value to the set of states in $U$ that eventually converge to the same state on $\chi$.
Suppose that ${\bm s}_{1}$ and ${\bm s}_{2}$ are taken from $U$, where ${\bm s}_{2}$ is on $\chi$ at phase $\theta$, i.e., ${\bm s}_2 = {\bm s}_0 (\theta)$. If ${\bm s}_1$ and ${\bm s}_2$ are asymptotically equivalent, we define the phase of ${\bm s}_1$ as
\begin{align}
	\Theta({\bm s}_1) = \Theta({\bm s}_2) = \theta. 	\label{eq. asphase}
\end{align}
Note that the convergence concept of the solutions in hybrid dynamical systems demands somewhat careful attention (see Appendix~\ref{sec. defphase} for details). 
Some properties of the isochron and the phase function on $U$ are discussed in Appendix~\ref{sec. isochron}.

The above definition of the phase guarantees that the following relation holds for almost all $t$ (excluding the Lebesgue measure zero set of the moments of switching) for an unperturbed oscillator:
\begin{align}
	\dot{\theta}(t) &= \dot{\Theta}(I(t), {\bm X}(t)) 
	= \nabla \Theta (I(t), {\bm X}(t)) \cdot {\bm F}(I(t),{\bm X}(t)) = 1, \label{eq. phasedef}
\end{align}
where $\nabla \Theta$ represents the gradient of $\Theta$ with respect to ${\bm X}$.
Namely, the phase $\theta$ rotates on a circle $\mathbb{T}^1$ at a constant frequency $1$.

When a sufficiently small perturbation $\epsilon {\bm p}(I(t), {\bm X}(t), t)$ with $|\epsilon| \ll 1$ is introduced to the oscillator as
\begin{align}
	\dot{{\bm X}}(t) =& {\bm F}(I(t), {\bm X}(t)) + \epsilon {\bm p}(I(t), {\bm X}(t), t), \label{eq. ptdphaseeq}
\end{align}
we can obtain the following approximate phase equation closed in $\theta$ at the lowest order:
\begin{align}
	\dot{\theta}(t) &= \dot{\Theta}(I(t), {\bm X}(t)) \cr
&= 1 \! + \! \epsilon \nabla \Theta (I_{0}(\theta), {\bm X}_0(\theta))\! \cdot \! {\bm p}(I_{0}(\theta), {\bm X}_0(\theta), t) \! + \! O (\epsilon^2 ) \cr
&= 1 + \epsilon {\bm Z}(\theta) \cdot {\bm p}(I_{0}(\theta), {\bm X}_0(\theta), t) + O (\epsilon^2 ),
\label{eq. reducedeq}
\end{align}
where we defined the phase sensitivity function 
\begin{align}
{\bm Z}(\theta) = \nabla \Theta (I_0(\theta), {\bm X}_0(\theta))
\end{align}
characterizing the linear response property of the oscillator phase to perturbations.
We consider that the phase evolves as a solution of a suitably regularized, multivalued system of Eq.~(\ref{eq. reducedeq}), such as the Filippov system~\cite{Filippov2013,Cortes2012discontinuous}.
(We do not consider impulsive perturbation at the moment of switching in this study, which requires special treatment.) 
The ideas underlying the phase approximation Eq.~(\ref{eq. reducedeq}) and some notes on the notion of the solution of it are given in Appendix~\ref{sec. approxphase}.

Thus, once we obtain the phase sensitivity function ${\bm Z}(\theta)$, the dynamics of a weakly perturbed hybrid limit cycle described by Eq.~(\ref{eq. ptdphaseeq}) can be reduced to a single phase equation~(\ref{eq. reducedeq}). 
Using the reduced phase equation, we can analyze various synchronization dynamics of hybrid limit cycles in detail. 
As we derive in Appendix~\ref{sec. derivation_adj}, ${\bm Z}(\theta)$ is given by a periodic solution to the following adjoint system:
\begin{align}
	&\dot{\bm Z}(t) = -{\bf A}^{\dag}(k,t){\bm Z}(t)  
	\mbox{\ \ for\ \ } t \ (\mathrm{mod}  \ T) \in (\tau_{k - 1}({\bm s}^*),\tau_{k}({\bm s}^*)), \\
	&{\bm Z}(t) = ({\bf C}_{k})^{\dag} {\bm Z}(t+0)  \mbox{\ \ at\ \ } t \ (\mathrm{mod} \ T) = \tau_{k}({\bm s}^*),
\end{align}
which is normalized to satisfy Eq.~(\ref{eq. phasedef}) on $\chi$, that is, 
\begin{align}
	{\bm Z}(t) \cdot {\bm F}(I_0(t), {\bm X}_0(t)) = 1.
	\label{eq. normalization}
\end{align}
Here, ${\bf A}(k,t) = D{\bm F}(k,{\bm X}_0(t))$ is the Jacobi matrix of ${\bm F}(k,{\bm X})$ estimated on $\chi$,
and ${\bf C}_k$ is a ``saltation matrix''~\cite{Bernardo08Piecewise} given by
\begin{align}
	\hspace*{-10mm}
	{\bf C}_k &= D {\bm \Phi}_k({\bm X}_0(\tau_k({\bm s}^*))) 
	\cr &
	\! - \!\left[ D {\bm \Phi}_k({\bm X}_0(\tau_k({\bm s}^*))) \dot{{\bm X}}_0(\tau_k({\bm s}^*))\! - \!\dot{{\bm X}}_0(\tau_k({\bm s}^*)\! + \! 0)\right]
	\! \otimes \!
	\left( \dfrac{\nabla L_k({\bm X}_0(\tau_k({\bm s}^*)))}{\nabla L_k({\bm X}_0(\tau_k({\bm s}^*)))\cdot \dot{{\bm X}}_0(\tau_k({\bm s}^*))} \right),
	\cr
	\label{eq. jumpmatrix} 
\end{align}
where $D {\bm \Phi}_k$ is the Jacobi matrix of ${\bm \Phi}_k$ and $\otimes$ represents a tensor product of two vectors.
${\bf C}_{k}$ represents expansion or contraction of small deviations from $\chi$ by the mapping ${\bm \Phi}_{k}$ at the switching $t=\tau_{k}({\bm s}^*)$, where the second term on the right-hand side takes into account the shift in the switching time caused by the perturbation.
In general, the above adjoint system can be integrated only backward in time because ${\bf C}_k$ can be singular.

In numerical calculations, we integrate these adjoint equations backward in time with occasional renormalization of ${\bm Z}(t)$ so Eq.~(\ref{eq. normalization}) is satisfied.
Then, reflecting the linear stability of $\chi$, all modes except the neutrally stable periodic solution decay and ${\bm Z}(\theta)$ is eventually obtained.
This is a standard procedure for calculating ${\bm Z}(\theta)$ of ordinary limit cycles and is called the adjoint method after Ermentrout~\cite{ermentrout2010mathematical}.

\section{Examples}

As an application of the phase reduction theory for hybrid limit cycles that we developed, we analyze injection locking of hybrid limit-cycle oscillators, i.e., synchronization of the oscillator to a periodic external signal~\cite{kuramoto2003chemical}.
We apply a weak periodic signal ${\bm p}(t) = {\bm p}(t + T_\mathrm{ext})$ to the hybrid limit cycle described by Eqs.~(\ref{eq. hybDS}) and (\ref{eq. resetF}). 
Using the phase reduction theory, the state of the perturbed oscillator, described by Eq.~(\ref{eq. ptdphaseeq}), can be approximately represented by its phase $\theta$, which obeys the reduced phase equation~(\ref{eq. reducedeq}). 

To analyze the synchronization dynamics, we consider the phase difference $\psi$ between the oscillator and the periodic signal,
\begin{align}
	\psi = \theta - \frac{T}{T_\mathrm{ext}}t,
	\label{eq. phasediff}
\end{align}
where $\theta$ is the phase of the hybrid limit cycle, $T$ is the natural period of the hybrid limit cycle, and $T_{\mathrm{ext}}$ is the period of the external periodic signal.
The frequency mismatch between the oscillator and the signal is given by $\epsilon \Delta = 1 - T/T_\mathrm{ext}$.
As shown in Appendix~\ref{sec. approxphasedyn}, using the standard averaging approximation for weakly perturbed oscillators~\cite{kuramoto2003chemical,hoppensteadt1997weakly,ermentrout2010mathematical}, the dynamics of $\psi$ can be derived from the reduced phase equation~(\ref{eq. reducedeq}) as
\begin{align}
	\dot{\psi} = \epsilon[\Delta + \Gamma(\psi)] \label{eq. averagedeq}
\end{align}
where the $T$-periodic phase coupling function $\Gamma(\psi)$ is given by
\begin{align}
	\Gamma(\psi) = \frac{1}{T_\mathrm{ext}}\int_0^{T_\mathrm{ext}}{{\bm Z}\left(\frac{T}{T_\mathrm{ext}}t + \psi\right)\cdot {\bm p}(t)}dt. \label{eq. convforpcf}
\end{align}
As mentioned previously for the phase equation, we consider that the phase difference evolves as a solution of Eq.~(\ref{eq. averagedeq}) in a regularized sense, if necessary. 
See~\cite{Perestyuk2013averaging+Perestyuk2011differential} for the averaging approximation in nonautonomous systems with jumps and multivalued righthand sides.

Synchronization dynamics of the oscillator can easily be understood from the phase coupling function $\Gamma(\psi)$. If Eq.~(\ref{eq. averagedeq}) has a stable fixed point, then the phase difference $\psi$ converges to this point and the oscillator is phase locked to the periodic signal. If there exist multiple stable fixed points, then the oscillator can be phase locked to the periodic signal at multiple phase differences depending on the initial condition. If Eq.~(\ref{eq. averagedeq}) does not have a fixed point, $\psi$ continues to increase or decrease and phase locking does not occur. 

\subsection{Glued Stuart-Landau oscillator}

As the first example, we introduce an analytically tractable model of a hybrid limit-cycle oscillator, which is constructed by gluing two Stuart-Landau oscillators (normal forms of the supercritical Hopf bifurcation~\cite{guckenheimer83nonlinear}) of different amplitudes.
The glued Stuart-Landau oscillator has two discrete states $I \in \{1, 2\}$ and a two-dimensional continuous state variable ${\bm X}(t) = (x(t),y(t))^{\dag}$, where $\dag$ denotes the transpose of a matrix. The dynamics is described by 
\begin{align}
	&{\bm F}(1,{\bm X})= \left(
		\begin{array}{c}
			x - ay - (x^2 + y^2)(x - by) \\
			ax + y - (x^2 + y^2)(bx + y)
		\end{array}
	\right), \nonumber \\ \\
	&{\bm F}(2,{\bm X})= \left(
		\begin{array}{c}
			x - ay - \alpha^2(x^2 + y^2)(x - by) \\
			ax + y - \alpha^2(x^2 + y^2)(bx + y)
		\end{array}
	\right), \nonumber \\ \\
	&{\bf \Phi}_1({\bm X}) = \left(
		\begin{array}{c}
		\dfrac{x}{\alpha} \\
		y
		\end{array}
		\right), \ 
	{\bf \Phi}_2({\bm X}) = \left(
		\begin{array}{c}
		\alpha x \\
		y
		\end{array}
		\right), \nonumber \\ \\
	&{\bf \Pi}_{1,2}=\{ {\bm X} \ | \ (L_1({\bm X})=0) \ \cap\ (x \le 0) \},
	\quad
	{\bf \Pi}_{2,1}=\{ {\bm X} \ | \ (L_2({\bm X})=0) \ \cap \ (x \ge 0) \}, \nonumber \\
	&L_1({\bm X}) = y, \ L_2({\bm X}) = -y,	
\end{align}
and the parameters are set as $a = 2\pi + 1,b=1$ and $\alpha = 2$.
With these parameter values, this system has a stable limit cycle of period $T=1$.
We take the origin of the phase $\theta=0$ at $I=1$ and ${\bm X}=(0,1)^{\dag}$, i.e., ${\bf \Theta}(1,(0,1)^{\dag})=0$.

\begin{figure*}[htbp]
	\includegraphics[width=\textwidth]{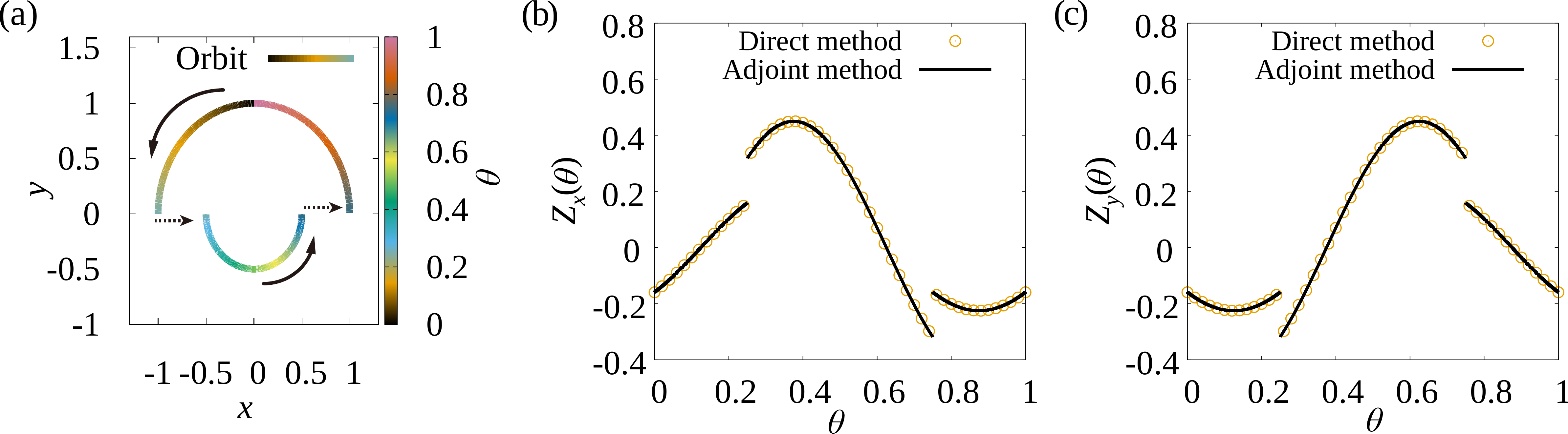}
	\caption{
	The glued Stuart-Landau oscillator. (a) The periodic orbit of a glued Stuart-Landau oscillator. The phase of the oscillator is shown in color code. The arrows represent the direction of the time evolution of the continuous state. The broken arrows indicate jumps. [(b) and (c)] The $x$ and $y$ components of the phase sensitivity function ${\bm Z}(\theta)$ obtained by the direct method (circles) and by the proposed adjoint method (lines).}
\label{fig. HSL}
\end{figure*}

The periodic orbit $\chi$ is depicted on $\mathbb{R}^2$~(Fig.~\ref{fig. HSL}(a)), which satisfies 
\begin{align}
(I_0(\theta),{\bm X}_0(\theta))=(1,(-\sin(2\pi \theta),\cos(2\pi \theta)))
\end{align}
for $\theta \in \mathrm{D}_1$, and
\begin{align}
(I_0(\theta),{\bm X}_0(\theta))=(2,(-0.5\sin(2\pi \theta),0.5\cos(2\pi \theta)))
\end{align}
for $\theta \in \mathrm{D}_2$, where $\mathrm{D}_1 = [0,0.25) \cup [0.75,1)$ and $\mathrm{D}_2 = [0.25, 0.75)$ are domains of the phase. 
			
The phase sensitivity function can be obtained by solving the adjoint linear problem analytically and is given by 
\begin{align}
{\bm Z}(\theta) = -\frac{1}{2\pi}(\cos{2\pi \theta}-\sin{2\pi \theta}, \sin{2\pi \theta} + \cos{2\pi \theta})^{\dag}
\end{align}
for $\theta \in \mathrm{D}_1$, and 
\begin{align}
{\bm Z}(\theta) = -\frac{1}{\pi}(\cos{2\pi \theta}-\sin{2\pi \theta}, \sin{2\pi \theta} + \cos{2\pi \theta})^{\dag}
\end{align}
for $\theta \in \mathrm{D}_2$. 

For the waveform of the periodic injection signal ${\bm p}(t)=(p_1(t), p_2(t))^\dag$, we consider rectangular waves:
\begin{align}
p_1(t) = -c \ (\mbox{if}\  t \ \mathrm{mod} \ T_\mathrm{ext} \in \mathrm{D}),
\quad
0 \ (\mbox{otherwise}),
\end{align}
and
\begin{align}
p_2(t)=0 \ \mbox{for all} \ t,
\end{align}
where the constant $c > 0$ is set so that the squared mean of the injection signal becomes unity, i.e., $\langle {\bm p}^2 \rangle \equiv \frac{1}{T_\mathrm{ext}}\int_0^{T_\mathrm{ext}}{{\bm p}^2(t)dt} =1$ unless otherwise specified, and $\mathrm{D}$ is the time domain where the forcing takes place. 

\begin{figure*}[htbp]
	\includegraphics[width=\textwidth]{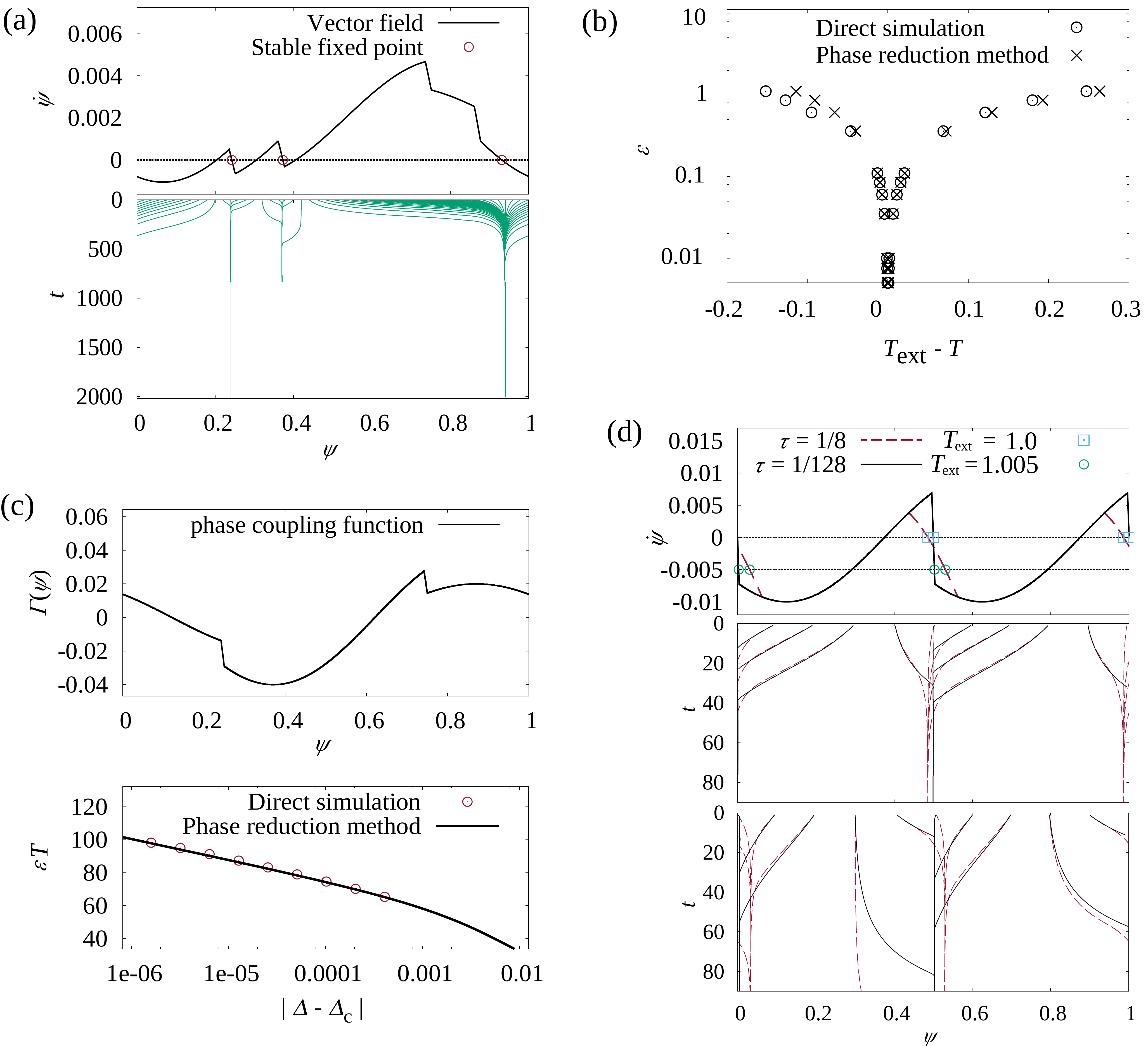}
	\caption{Phase reduction analysis of the injection locking of a glued Stuart-Landau oscillator.
	(a)
	Dynamics of the phase difference $\psi$. 
	Time derivative $\dot{\psi}$ plotted as a function of $\psi$, where three stable fixed points (circles) coexist (top panel).
	Trajectories of $\psi$ from 50 different initial states obtained by direct numerical simulation of the original model, converging to the stable fixed points (bottom panel).
	The parameters are set as $\epsilon = 0.1$, $T_\mathrm{ext}/T = 1.002$, and the domain where $p_1(t)$ takes a nonzero value is $\mathrm{D} = \{t \ | \ 0 \le t \le T_\mathrm{ext}/64 \ \cup \ 3T_\mathrm{ext}/8 \le t \le 25T_\mathrm{ext}/64  \}$.
	(b) The Arnold tongue showing the region where phase locking takes place. Here $\mathrm{D} = \{t \ | \ 0 \le t \le T_\mathrm{ext}/2 \}$. 
(c) The phase coupling function with sharp corners (top panel) and the period of the phase slipping plotted in log-linear scales (bottom panel). Here $\epsilon =0.01$ and $\mathrm{D} = \{t \ | \ 0 \le t \le T_\mathrm{ext}/128 \}$. } 
	\label{fig. pHSL}
\end{figure*}
\addtocounter{figure}{-1}
\begin{figure*}[htbp]
	\caption{(continued) 
(d) Dynamics of the phase difference $\psi$ for the mild, nonimpulsive (yellow line) and impulsive (black line) signals. Time derivative $\dot{\psi}$ vs. $\psi$ with stable fixed points (circles) for the different frequencies of the input for each case (top panel). Trajectories of $\psi$ from 10 different initial states for $T_\mathrm{ext}/T=1.0$ (middle panel) and for $T_\mathrm{ext}/T=1.005$ (bottom panel). The parameters are set as follows: $\epsilon = 0.01$, and $\mathrm{D} = \{t \ | \ | t - T_\mathrm{ext}/4|\le \tau T_\mathrm{ext}/2 \ \cup \ | t - 3T_\mathrm{ext}/4|\le \tau T_\mathrm{ext}/2\}$. 
}
\end{figure*}

In Figs.~\ref{fig. HSL}(b) and~\ref{fig. HSL}(c), the phase sensitivity function ${\bm Z}(\theta)$ obtained by analytically solving the proposed adjoint systems is compared with the result of the direct numerical simulation (see Appendix~\ref{sec. directmethod} for details). The results are in good agreement and show the validity of the proposed adjoint method. The discontinuities in ${\bm Z}(\theta)$ are characteristic of a hybrid limit-cycle oscillator.

Figure~\ref{fig. pHSL}(a) displays the averaged dynamics of the phase difference $\psi$ for several initial values, where the result of phase reduction is compared with direct numerical simulations. It can be seen that the asymptotic phase differences and their dependence on initial conditions are well predicted from the phase coupling function $\Gamma(\psi)$. 
Figure~\ref{fig. pHSL}(b) shows the boundaries of the region where the injection locking takes place, called the Arnold tongue~\cite{hoppensteadt1997weakly}. Results of the numerical simulation also agree well with the analytical prediction by the phase reduction theory. 
Thus, the injection locking of hybrid limit-cycle oscillators by weak periodic input can be theoretically predicted by using ${\bm Z}(\theta)$ obtained by the adjoint method.

Here, we emphasize one peculiar property of the hybrid oscillator.
For smooth oscillators, the period of the phase slipping near the critical point generally obeys the inverse square-root scaling law, $\epsilon T_{\mathrm{slip}} \sim |\Delta - \Delta_c|^{-1/2}$, where $\Delta_c$ is the critical value of $\Delta$~\cite{kuramoto2003chemical,izhikevich2010dynamical}, because the phase coupling function $\Gamma(\psi)$ generally has quadratic maximum and minimum.
In contrast, as shown in Fig.~\ref{fig. pHSL}(c), 
the hybrid oscillator with discontinuous ${\bm Z}(\theta)$ can possess nonsmooth $\Gamma(\psi)$ with sharp nonquadratic maximum or minimum when subjected to impulsive signals. 
For such $\Gamma(\psi)$, we can show that the period of phase slipping $T_{\mathrm{slip}}$ obeys
\begin{align}
	\epsilon T_{\mathrm{slip}} \sim - \ln{|\Delta-\Delta_c|} 
\end{align}
at the leading order in the vicinity of the critical value $\Delta_c$, where $\Delta_c+\Gamma (\psi^*)=0$ and $\Gamma(\psi^*)$ is the extremum at the corner of $\Gamma (\psi)$, and the semiderivatives $\Gamma' (\psi^*-0)$ and $\Gamma'(\psi^*+0)$ are nonzero (see Appendix~\ref{sec. neglog} for the derivation).
Since nonsmooth corners in $\Gamma (\psi )$ cannot exist in smooth systems, singular scaling law of this kind is characteristic of hybrid oscillators.

In Fig.~\ref{fig. pHSL}(d), the transient dynamics of $\psi$ for two different types of rectangular wave input, one with a low-duty ratio (impulsive) $\tau=1/128$ and the other with a mediate one (mild, nonimpulsive) $\tau=1/8$, is compared. Here the magnitude $c$ of the input signal is normalized so that the uniform norm of $\Gamma$ becomes unity, i.e., $\max_\psi{|\Gamma(\psi)|}=1$, for each case. 
For the impulsive case, $\psi$ approaches the stable phase difference $\psi_0$ with a nonzero angle, while in the nonimpulsive case, the approach is tangential.
This implies that the decay of the deviation from $\psi_0$ is faster than exponential in the impulsive case and the time required to establish entrainment is drastically shorter.
Moreover, variations in the input period only slightly changes $\psi_0$ for the impulsive input, while $\psi_0$ shows significant change for the nonimpulsive input.
This ultrafast entrainment and robustness of the stable phase difference can be attributed to the existence of the region $\mathrm{D}_{\mathrm{s}}$ where $\Gamma (\psi)$ is extremely steep.

Note that the discontinuity in ${\bm Z}(\theta)$ is necessary for the existence of such a region $\mathrm{D}_{\mathrm{s}}$,
because $\Gamma(\psi)$ is given by the convolution~(\ref{eq. convforpcf}) of ${\bm Z}(\theta)$ and ${\bm p}(t)$; when the input is an ideal impulse, the slope of $\Gamma(\psi)$ in $\mathrm{D}_{\mathrm{s}}$ can be infinite.  
Therefore, these interesting synchronization properties are distinctive feature of the hybrid oscillators driven by impulsive periodic input.
Such a type of very fast (or finite-time) synchronization has been studied for simple neuron models with discontinuity whose ${\bm Z}(\theta)$ can be obtained analytically, as well as in some fast-slow models in the fast-relaxation limit~\cite{izhikevich2000phase,veryfast,veryfast2}.
Our argument based on the phase reduction theory for hybrid limit cycles is general and can be applied to high-dimensional systems where the nonsmoothness is not the result of adiabatic approximation. 

\subsection{Passive bipedal walker}

\begin{figure*}
	\includegraphics[width=\textwidth]{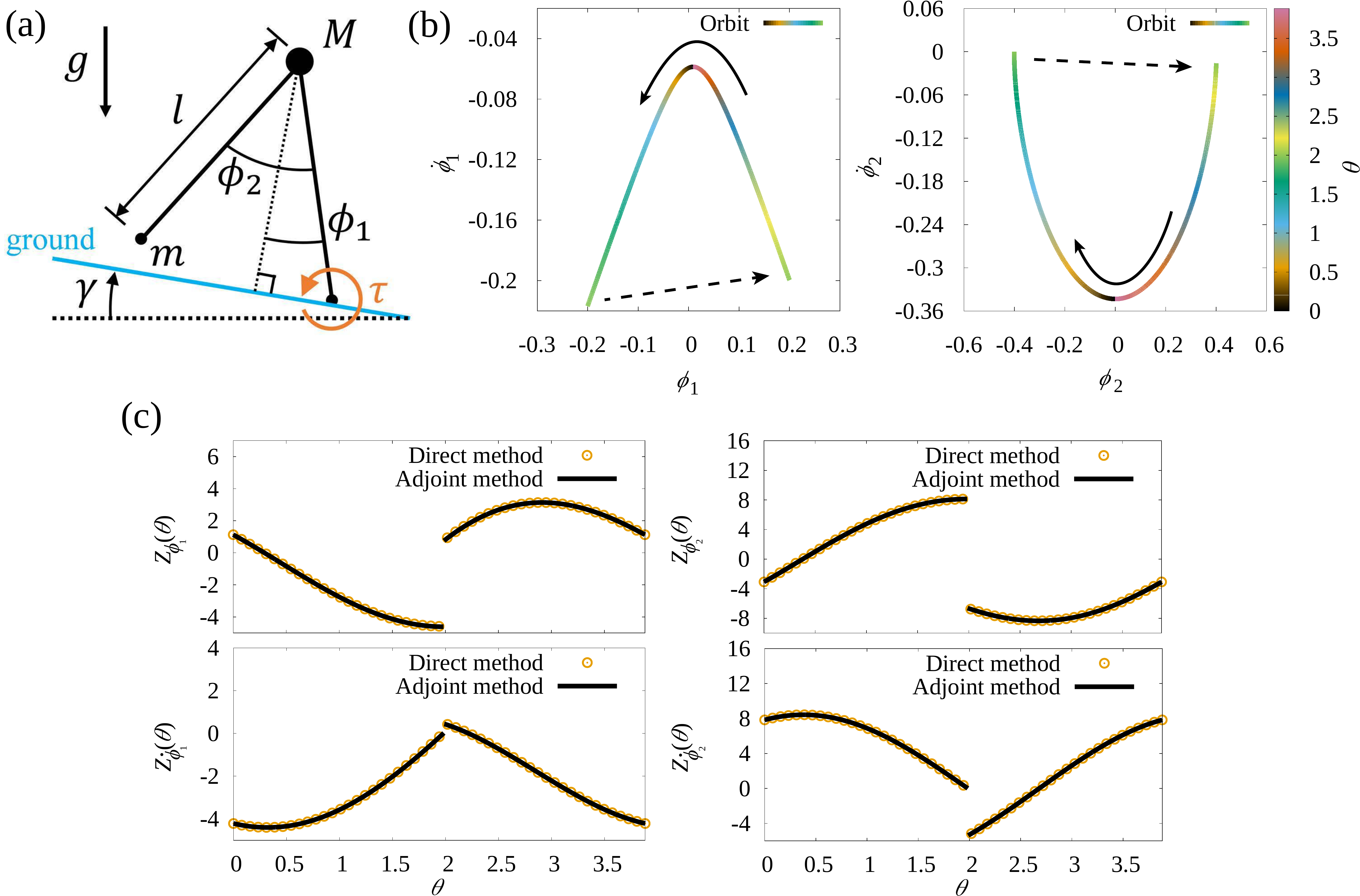}
	\caption{
	Two-link model of a passive walker walking down a slope. (a) Schematic of the model. (b) The periodic orbit of the model. The orbit discontinuously jumps when the swing and support leg alternate with each other. The arrows represent the direction of the time evolution of the continuous state. The broken arrows indicate jumps. The phase is shown in color code. 
	(c) Four components of the phase sensitivity function ${\bm Z}(\theta) = (Z_{\phi_1}(\theta), Z_{\dot{\phi}_1}(\theta), Z_{\phi_2}(\theta), Z_{\dot{\phi}_2}(\theta))^{\dag}$ obtained by the direct method (circles) and by the proposed adjoint method (lines).}
\label{fig. passive}
\end{figure*}

Next, we analyze a physical example of hybrid limit-cycle oscillator, namely a two-link model of a passive walker walking down a slope~\cite{garcia1998simplest}, proposed as a simple model of biped locomotion. Figure~\ref{fig. passive}(a) shows a schematic diagram of the model, where $g$ is the gravitational acceleration; $l$ is the length of the legs; $M$ and $m$ are the masses of the hip and the foot, respectively; $\phi_1$ and $\phi_2$ specify the angles of the swing and support legs; $\gamma$ is the angle of the slope; and $\tau$ is a periodic torque applied to the ankle of the support leg. 
It is assumed that $m/M = 0$, i.e., the hip mass is much larger than the foot mass, the tip of the support leg does not slip along the ground, and the collision of the foot with the ground is perfectly inelastic (no slip and no bounce). 
This model exhibits a stable limit-cycle oscillation for appropriate parameter values that corresponds to periodic movements of the legs. 
This is a four dimensional hybrid dynamical system with impacts, hence it cannot be dealt with by the conventional methods~\cite{coombes2012nonsmooth,izhikevich2000phase,park2013infinitesimal} mentioned above. 

The model has a one discrete state $I \in \{1\}$ and a continuous state variable ${\bm X}(t) = (\phi_1(t),\dot{\phi}_1(t),\phi_2(t),\dot{\phi}_2(t))^{\dag}$. 
The dynamics is described by 
\begin{align}
			&{\bm F}(1,{\bm X}) 
			= \left(
		\begin{array}{c}
			\dot{\phi}_1 \\
			\sin{(\phi_1 - \gamma)} \\
			\dot{\phi}_2 \\
			\sin{(\phi_1 - \gamma)} + \dot{\phi}_1^2 \sin{\phi_2} - \cos{(\phi_1 - \gamma)}\sin{\phi_2}
		\end{array}
	\right), \nonumber \label{eq. eqm} \\ \\
	& {\bf \Phi}_1({\bm X}) = \left(
		\begin{array}{c}
			-\phi_1 \\
			\dot{\phi}_1 \cos{2\phi_1} \\
			-2\phi_1 \\
			\dot{\phi}_1 \cos{2\phi_1}(1 - \cos{2\phi_1})
		\end{array}
	\right), \nonumber \label{eq. coll} \\ \\
	&{\bf \Pi}_{1,1} = \{ {\bm X} \ | \ (L((1,1),{\bm X})=0) \cap (\phi_2 < -\delta) \}, \nonumber \\ 	&L_1({\bm X}) = 2\phi_1 - \phi_2, \label{eq. collc}
\end{align}
where we have rescaled time by $\sqrt{l/g}$, and $\delta > 0$ is a small positive constant (we set $\delta = 0.1$), which is introduced to avoid foot scuffing (contact of the swing leg with the ground in the middle of the swing).
The parameter $\gamma$ is set as $\gamma = 0.009$. 
Note that Eq.~(\ref{eq. eqm}) is an equation of motion representing continuous dynamics of the walker during the single-leg support phase, in which the walker stands on the support leg and moves the swing leg, where $\phi_1$ and $\phi_2$ are the angular coordinates of the swing and support legs. See Ref.~\cite{goswami1996compass} for a detailed derivation of the above type of equations in nearly the same setting. 
Note also that the main effect of the inclined ground to the walking dynamics, i.e., the ground reaction force, is already included in the dynamical model. This effect is not considered a perturbation and therefore need not be weak, as long as the model exhibits stable rhythmic walking. If there exist additional small effects from the flat inclined ground, such as slight up and down, they can be incorporated into the reduced phase model perturbatively. 
Finally, although the motion of Eq.~(\ref{eq. eqm}) during the single-leg support phase appears to be conservative, the collision of the leg with the ground, described Eq.~(\ref{eq. coll}) and Eq.~(\ref{eq. collc}), is perfectly inelastic (plastic), so the impact of the leg with nonzero velocity relative to the ground causes energy dissipation. This energy loss is compensated by the gravitational potential energy, which is supplied to the system at each moment of the collision of the leg with the ground. Thus, a stable limit cycle can arise in this hybrid dynamical system. 

Figure~\ref{fig. passive}(b) shows the stable periodic orbit of the model. 
Using the shooting method developed in~\cite{khan2011sensitivity}, a point on $\chi$, which we define as the origin of the phase, and the period $T$ of the orbit can be obtained as
\begin{align}
{\bm s}^* = (1,(0.009000,-0.05869,-0.0009629,-0.3432)^{\dag}), \quad T=3.882.
\end{align}

\begin{figure*}
	\includegraphics[width=\textwidth]{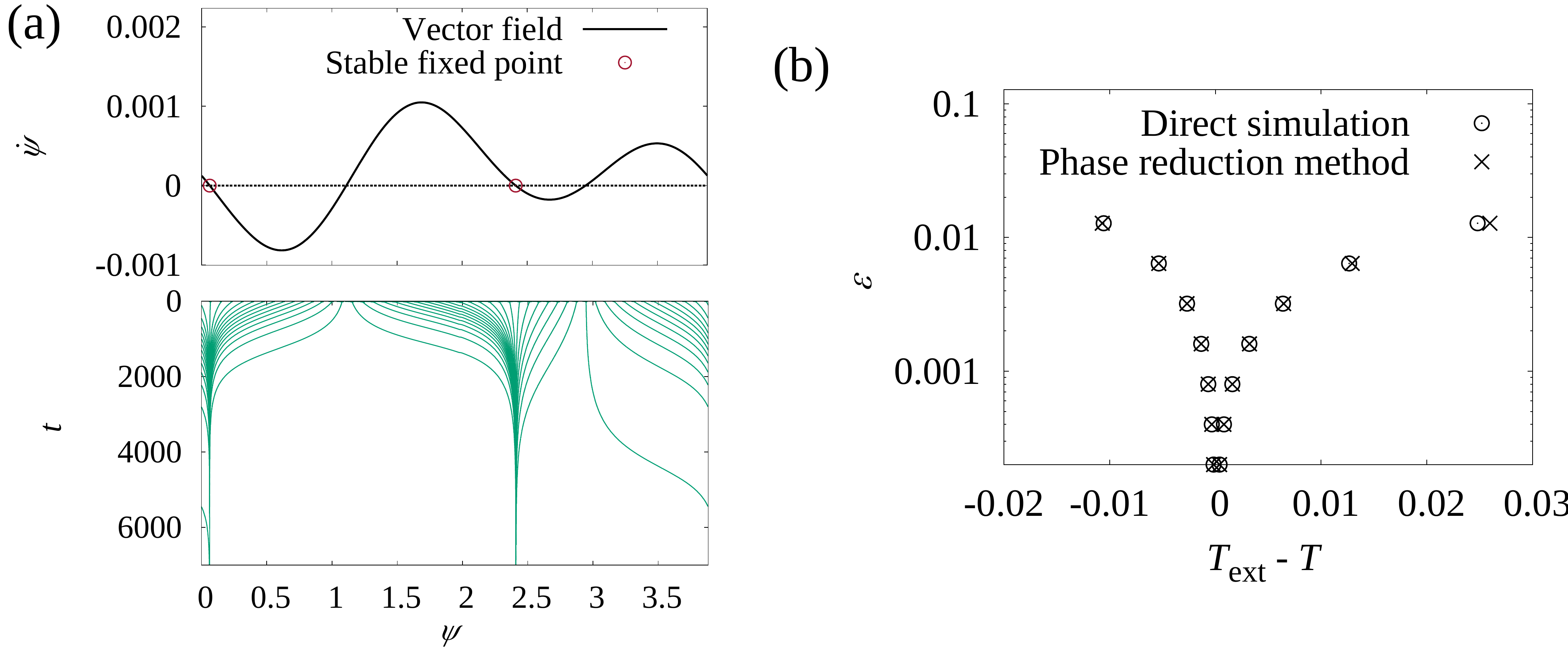}
	\caption{
	Phase reduction analysis of the injection locking of the two-link model of the passive walker. (a) Dynamics of the phase difference $\psi$ (line). $\dot{\psi}$ vs. $\psi$, where stable fixed points are represented by circles (top panel). Trajectories of $\psi$ from 50 different initial states obtained by direct numerical simulation of the original model (bottom panel). The parameters are set as $\epsilon = 0.00253$ and $T_\mathrm{ext}/T = 1.0005$. (b) The Arnold tongue obtained by the phase reduction and by direct numerical simulation of the original model.}
\label{fig. ppassive}
\end{figure*}

Figure~\ref{fig. passive}(c) shows the phase sensitivity function ${\bm Z}(\theta) = (Z_{\phi_1}(\theta), Z_{\dot{\phi}_1}(\theta), Z_{\phi_2}(\theta), Z_{\dot{\phi}_2}(\theta))^{\dag}$ with discontinuity in the middle, which is obtained numerically by the proposed adjoint method. The result agrees well with the one obtained by the direct method, thus the proposed adjoint method also works nicely for this model.

Using the reduced phase equation, we study the injection locking of the passive walker to the periodic ankle torque actuation. 
That is, we apply a weak periodic torque to the walker, where the frequency of the torque is close to that of the natural frequency of the walker, and analyze whether the walker synchronizes with the applied weak torque. 
We introduce periodic actuation of the ankle torque as the injection signal ${\bm p}(t)=(0,\tau (t),0,0)^{\dag}$, where
\begin{align}
\tau (t) = c e^{-0.5[t \ (\mathrm{mod} \ T_\mathrm{ext})]/T_\mathrm{ext}}\sin{(4\pi t/T_\mathrm{ext})}.
\end{align}
The magnitude $c$ of the waveform is determined to satisfy the normalization condition $\langle {\bm p}^{2} \rangle = 1$.
As in the case of the glued Stuart-Landau oscillator, we can obtain the phase coupling function $\Gamma(\psi)$ from the phase sensitivity function ${\bm Z}(\theta)$ and the injected periodic signal ${\bm p}(t)$, and predict the dynamics of the phase difference $\psi$ between the oscillator and the signal. 

Figure~\ref{fig. ppassive}(a) shows the dynamics of $\psi$. The reduced phase equation predicts that there are two stable fixed points of $\psi$, and direct numerical simulations of the original model from different initial conditions confirm that the phase difference of the passive walker is actually attracted to either of the stable fixed points.
Figure~\ref{fig. ppassive}(b) plots the Arnold tongue showing the region where the phase locking of the passive walker to the injected signal takes place. The results obtained by the phase reduction theory agree well with the results obtained by direct numerical simulations of the original model. 

Thus, the proposed phase reduction theory is also useful in analyzing realistic physical systems, even when the hybrid limit-cycle oscillator has a high-dimensional continuous state.

\section{Summary}

We formulated a phase reduction theory for a general class of hybrid limit-cycle oscillators and derived the adjoint equation for the phase sensitivity function.
The proposed theory provides precise phase sensitivity functions and the derived phase equation accurately predicts the injection locking properties of hybrid oscillators.
We illustrated synchronization properties characteristic to hybrid oscillators, such as ultrafast entrainment to periodic signal and negative logarithmic scaling at the synchronization transition, and explained them by using the reduced phase equation with discontinuous phase sensitivity functions.

The phase reduction theory developed in this study would serve as a powerful tool for investigating synchronization phenomena in complex nonsmooth systems and for finding various applications in controlling distributed interacting nonlinear oscillators~\cite{Ijspeert98central+Strogatz05,Dunnmon11power+Moss10low,tanaka2009self}.

\begin{acknowledgments}

We acknowledge the valuable comments on this work of Jaap Eldering, Shigefumi Hata and Hiroshi Kokubu.
This study was supported by Japan Society for the Promotion of Science (Japan) KAKENHI (Grants No.~22684020, No.~25540108, No.~26103510, No.~26120513 and No.~15J12045), the Core Research for Evolutional Science and Technology (Japan) Kokubu project of Japan Science and Technology Agency (Japan), and the FIRST Aihara project of Japan Society for the Promotion of Science (Japan). 

\end{acknowledgments}

\appendix

\section{Assumptions for the periodic solution
\label{sec. assumptions}}

In this section, we introduce the assumptions that are necessary for the periodic solution to be piecewise continuously differentiable with respect to the initial condition~\cite{Akhmet2005on+akhmet2010principles}.
Hybrid dynamical systems can exhibit pathological behaviors, which do not occur in smooth dynamical systems, such as the grazing, livelock, sliding, and Zeno phenomena due to the effect of discrete switching~\cite{Bernardo08Piecewise,Schaft00}. 
The grazing phenomenon~\cite{Bernardo08Piecewise} occurs when the orbit becomes tangent to the switching surface. This condition can be written as 
\begin{align}
	\nabla L((i,j),{\bm X}_0(t))|_{L=0} \cdot \dot{{\bm X}}_0(t)=0,
\end{align}
where $\nabla L:\mathcal{G} \times \mathbb{R}^{N} \to \mathbb{R}^{N}$ is the gradient of $L$ with respect to ${\bm X}$ and $\cdot$ denotes inner product of vectors. 
The livelock, sliding, and Zeno phenomena~\cite{Schaft00} can arise when 
the points ${\bm X}_0(\tau_k({\bm s}^*)+0), \mbox{\ }k\in \mathbb{Z}_{\ge 0}$ are allowed to be accumulation points of the switching surfaces.
These conditions can lead to infinite sensitivity to the initial conditions~\cite{jeffrey2011nondeterminism}.
In this study, we do not consider such pathological situations, namely, we assume that (C1) the orbit is always transversal to the switching plane and that (C2) each continuous state right after the discrete state transition has a neighborhood that is disjoint from the switching surfaces.

\section{Linear stability of the hybrid limit cycle \label{sec. linearstability}}

In this section, we formalize the linear stability of the periodic solution. 
Let ${\bm \xi}^{\alpha}$ ($\alpha=1,...,N$) be the $\alpha$th initial-condition sensitivity vector~\cite{khan2011sensitivity} with respect to an initial state ${\bm s}^* = (I_{0}(0), {\bm X}_{0}(0))$ on $\chi$ at $t=0$, defined as
\begin{align}
{\bm \xi}^{\alpha}(t) \! = \! \lim_{\epsilon \to 0}{ \left( \frac{ {\bm X} (t ; (I_{0}(0), {\bm X}_0(0)+\epsilon {\bm e}^{\alpha}))\! -\! {\bm X}_0(t)}{ \epsilon} \right)}.
\label{eq. defvariation}
\end{align}
Here, the second argument of ${\bm X}(t;\cdot)$ represents an initial state that is slightly perturbed in the $\alpha$th direction (${\bm e}^{\alpha}$ is the $\alpha$th unit vector) from $(I_{0}(0), {\bm X}_{0}(0))$ on $\chi$.
We introduce a {\it sensitivity matrix} ${\bf \Xi} = ({\bm \xi}^1, {\bm \xi}^2, \cdots ,{\bm \xi}^N) \in \mathbb{R}^{N\times N}$ as the collection of sensitivity vectors in all directions.  Note that
\begin{align}
	&{\bf \Xi}(0) = ({\bm e}^{1}, ..., {\bm e}^{N}) = {\bf I} \label{eq. vinit}
\end{align}
where ${\bf I}$ is the identity matrix, because ${\bm X} (0 ; (I_{0}(0), {\bm X}_0(0)+\epsilon {\bm e}^{\alpha})) = {\bm X}_{0}(0) + \epsilon {\bm e}^{\alpha}$. 

In Refs.~\cite{Akhmet2005on+akhmet2010principles,khan2011sensitivity}, the linear variational system for ${\bf \Xi}$ has been derived as
\begin{align}
	\dot{{\bf \Xi}}(t) = {\bf A}(k,t) {\bf \Xi}(t)
	\mbox{\ \ for\ \ } t \ (\mathrm{mod} \ T)\in (\tau_{k-1}({\bm s}^*),\tau_{k}({\bm s}^*)), \label{eq. vareqsmooth}
\end{align}
\begin{align}	
	&{\bf \Xi}(t+0) = {\bf C}_k {\bf \Xi}(t) \mbox{\ \ at\ \ } t \ (\mathrm{mod} \ T) =\tau_{k}({\bm s}^*), \label{eq. varjump} 
\end{align}
where ${\bf A}(k,t) = D{\bm F}(k,{\bm X}_0(t))$ is the Jacobi matrix of ${\bm F}(k,{\bm X})$ estimated on $\chi$, and ${\bf C}_k$ is the so-called {\it saltation matrix}~\cite{Bernardo08Piecewise} (Eq.~(\ref{eq. jumpmatrix}) in the main article).

We define a {\it monodromy matrix} ${\bf M}$ from the sensitivity matrix ${\bf \Xi}(t)$  as ${\bf M} = {\bf \Xi}(T)$.
If this ${\bf M}$ has one simple eigenvalue equal to 1 and all other eigenvalues lie strictly inside the unit circle on the complex plane, the periodic solution is linearly stable~\cite{khan2011sensitivity}.
We call such a stable isolated periodic solution a hybrid limit cycle. 
It can be shown that the eigenvalues and their algebraic multiplicities of monodromy matrices do not depend on the choice of the initial state. 
We can also consider initial-condition sensitivity vectors with respect to the state ${\bm s}_0(\theta)=(I_{0}(\theta), {\bm X}_{0}(\theta))$ on the limit cycle $\chi$, instead of ${\bm s}^*=(I_{0}(0), {\bm X}_{0}(0))$, as
\begin{align}
&{\bm \xi}^{\alpha}(t;\theta) 
= \lim_{\epsilon \to 0}{\! \left( \! \frac{ {\bm X} (t ; (I_{0}(\theta), {\bm X}_0(\theta)\!+\!\epsilon {\bm e}^{\alpha}))\! -\! {\bm X}_0(t \! + \! \theta)}{ \epsilon}\! \right)}.
\label{eq. defvariation app}
\end{align}
We denote by a matrix ${\bf \Xi}(t;\theta) = ({\bm \xi}^1(t;\theta), {\bm \xi}^2(t;\theta), \cdots ,{\bm \xi}^N(t;\theta)) \in \mathbb{R}^{N \times N}$ the collection of the sensitivity vectors in all directions, and introduce a monodromy matrix ${\bf M}(\theta) = {\bf \Xi}(T;\theta)$ of the linear variational system, which satisfies 
\begin{align}
	&\dot{{\bf \Xi}}(t;\theta) = {\bf A}(k,t+\theta) {\bf \Xi}(t;\theta)
	\mbox{\ \ for\ \ } t + \theta \ (\mathrm{mod} \ T)\in (\tau_{k-1}({\bm s}^*),\tau_{k}({\bm s}^*)),  \label{eq. varx}
\end{align}
\begin{align}	
	&{\bf \Xi}(t\! + \! 0;\theta)\! = \! {\bf C}_k {\bf \Xi}(t;\theta) \mbox{\ \ at\ \ } t \! + \! \theta \ (\mathrm{mod} \ T) \! = \! \tau_{k}({\bm s}^*),  \label{eq. varx2}
\end{align}
\begin{align}
	{\bf \Xi}(0;\theta) = {\bf I}.
\end{align}

There exist a unique state transition matrix ${\bf H}_k(t,s)$
of the linear variational system (\ref{eq. vareqsmooth})-(\ref{eq. varjump}) satisfying
\begin{align}
	&{\bf \Xi}(t;\theta)={\bf H}_k(t+\theta,s+\theta) \ {\bf \Xi}(s;\theta)
	\mbox{\ \ for\ \ }	t+\theta, s+\theta \in [\tau_{k-1}({\bm s}^*)+nT,\tau_{k}({\bm s}^*)+nT),
	\label{eq. transfer}
\end{align}
which is a solution to
\begin{align}
	&\dot{{\bf H}}_k(t,s) ={\bf A}(k,t) \ {\bf H}_k(t,s)
	\mbox{\ \ for\ \ }	t, s \in [\tau_{k-1}({\bm s}^*)+nT,\tau_{k}({\bm s}^*)+nT], \label{eq. odetransfer}
\end{align}
with the initial condition ${\bf H}_k(s,s) = {\bf I}$.

Suppose that $\theta$ is in the interval $[\tau_{m^*-1}({\bm s}^*),\tau_{m^*}({\bm s}^*))$, where $m^{*} \in \mathcal{M}_{0}\cup \{ m_0 + 1\}$.
Using ${\bf H}_k(t,s)$, the monodromy matrix can be expressed as ${\bf M}(0) = {\bf M}_{1}{\bf M}_{2}$, where
\begin{align}
	{\bf M}_{1}={\bf H}_{1}(T,\tau_{m_0}({\bm s}^*))&\left(\ \sideset{}{^+}\prod_{l=m^*}^{m_0}{{\bf H}_{l+1}(\tau_{l+1}({\bm s}^*),\tau_{l}({\bm s}^*)){\bf C}_l}\right) 
	\cdot {\bf H}_{m^*}(\tau_{m^*}({\bm s}^*),\theta),
\end{align}
and
\begin{align}
	{\bf M}_{2} &=  {\bf H}_{m^*}(\theta,\tau_{m^*-1}({\bm s}^*)) 
	\cdot \sideset{}{^+}\prod_{k=1}^{m^*-1}{{\bf C}_k {\bf H}_k(\tau_k({\bm s}^*),\tau_{k-1}({\bm s}^*))}. 
\end{align}
Here, we denote by $\sideset{}{^+}\prod_{i=1}^{n}{{\bf Y}_i} = {\bf Y}_n {\bf Y}_{n-1} \cdots {\bf Y}_2 {\bf Y}_1$ the ordered product of matrices.
With these matrices ${\bf M}_{1}$ and ${\bf M}_{2}$,
one can also show that ${\bf M}(\theta) = {\bf M}_{2} {\bf M}_{1}$, where 
${\bf H}_k(t,s) = {\bf H}_k(t+nT,s+nT)$, which follows from ${\bf A}(k,t) = {\bf A}(k,t+nT)$,
is used.
Therefore, ${\bf M}(0)$ and ${\bf M}(\theta)$ has the same set of eigenvalues with the same algebraic multiplicities, because the Jordan blocks with nonzero eigenvalues of the products of the matrices ${\bf AB}$ and ${\bf BA}$ are identical~\cite[Th.~3.2.11.1.]{horn2012}. 

\section{Asymptotic equivalence of initial conditions in hybrid dynamical systems
\label{sec. defphase}}

In this section, we introduce an asymptotic equivalence of the initial conditions that we use for defining the isochrons, which differs from the one used in smooth systems. 
Suppose ${\bm X}_1(t)$ and ${\bm X}_2(t)$ are the continuous parts of the solutions to a system with initial conditions ${\bm s}_1$ and ${\bm s}_2$, respectively. 
In smooth systems, the asymptotic equivalence relation of the initial conditions ${\bm s}_1$ and ${\bm s}_2$ is defined as the convergence of the error ${\bm X}_1(t)-{\bm X}_2(t)$ in the Euclidean topology. That is, if $\lim_{t \to +\infty}{|{\bm X}_1(t)-{\bm X}_2(t)|}=0$ where $|\cdot |$ is the Euclidean norm, then ${\bm s}_1$ and ${\bm s}_2$ are asymptotically equivalent. 
In hybrid dynamical systems, the moments of switching of the two solutions for these initial conditions generally do not coincide in a finite time. 
Hence, the Euclidean norm of the error of the continuous part $|{\bm X}_1(t) - {\bm X}_2(t)|$ causes some kind of ``peaking behavior''~\cite{biemond2013tracking}; $|{\bm X}_1(t) - {\bm X}_2(t)|$ can be larger than a constant $c > 0$ (of order $|{\bm \Phi}_k({\bm X}) - {\bm X}|$) for some $t \in [t_*, \infty]$ for an arbitrarily large $t_* > 0$ due to the continuous state jumps. 
This violates the definition of convergence in Euclidean topology. Therefore, we need to consider convergence in some other suitable topology to define the asymptotic equivalence notion in hybrid dynamical systems.

Various topologies suitable for hybrid dynamical systems have been proposed in the literature, such as the Skorohod topology~\cite{broucke2002continuous} originally designed as a tool to analyze stochastic processes, the topology of graphical convergence that is based on set-valued analysis~\cite{goebel2006solutions}, and the quotient topology generated on the hybrifold~\cite{simic2005towards}, which is, roughly speaking, the manifold constructed by identifying the switching surface ${\bm \Pi}_{k,k+1}$ with its image of the transition function ${\bm \Phi}_k$. 

In this study, we adopt an asymptotic equivalence that is based on convergence in B-topology~\cite{Akhmet2005on+akhmet2010principles}, which is defined as follows: 
if for any $\epsilon > 0$, there exist $T^* = T^*(\epsilon) >0$ such that, 
in the time domain $[T^*,+\infty)$, every moment of switching of the solution ${\bm X}_2(t)$ lies in some $\epsilon$ neighborhood of the moment of switching of the solution ${\bm X}_1(t)$, 
and for all $t \in [T^*, +\infty)$, which are outside the $\epsilon$ neighborhoods of the moments of switching of ${\bm X}_1(t)$, $|{\bm X}_1(t)-{\bm X}_2(t)|<\epsilon$ holds, 
then we call the initial conditions ${\bm s}_1$ and ${\bm s}_2$ asymptotically equivalent. 
The benefit of this definition is intuitively clear;
the error $|{\bm X}_1(t) - {\bm X}_2(t)|$ is evaluated outside of the neighborhoods of the points of discontinuity,
and thus we can avoid the effect of the peaking behavior,
and the error that occurs at the moment of switching is guaranteed to disappear.

\section{Some properties of the isochron and the phase function
\label{sec. isochron}}

Using the following equivalence relation (Eq.~$(6)$ in the main article)
\begin{align}
	\Theta({\bm s}_1) = \Theta({\bm s}_2) = \theta,
	\label{eq. asphase app}
\end{align}
we can introduce the ``conditional'' isochron $W_k (\theta) = \{{\bm X} \: | \: \Theta(k,{\bm X})= \theta \}$, i.e., the set of continuous states sharing the same phase value $\theta$
for each discrete state $k \in {\mathcal M}_0$.
We can then define the isochron of $\chi$ with phase $\theta$ as the union $W(\theta) = \bigcup_{k=1}^{m_0}{(k,W_k(\theta))}$.
We note that the notion of the isochron in hybrid dynamical systems has been proposed in~\cite{burden2014hybrid}, but the phase dynamics of weakly perturbed hybrid oscillators has not been discussed so far.

We can show that, in a domain $\tilde{U} \equiv \mathcal{A} \setminus \bigcup_{k=1}^{m_0}{(k,{\bm \Pi}_{k,k+1})}$, where $\mathcal{A} \subset U$ is a neighborhood of $\chi$ such that the solution starting from the point in $\mathcal{A}$ is piecewise continuously diffentiable with respect to the initial condition, the phase function $\Theta$ is totally differentiable with respect to the continuous state and that it is one-sided differentiable on the switching surfaces as follows.

We consider two slightly different initial conditions, ${\bm s}_1$ and ${\bm s}_2 = {\bm s}_1 + (0,\epsilon {\bm h})$, where ${\bm s}_1, {\bm s}_2 \in \mathcal{A}$, $0<\epsilon \ll 1$ and ${\bm h}\in \mathbb{R}^N$. Note that, when ${\bm s}_1 \in \bigcup_{k=1}^{m_0}{(k,{\bm \Pi}_{k,k+1})}$, ${\bm h}$ should be taken from the subset of $\mathbb{R}^N$ whose elements point in the opposite direction as ${\bm F}({\bm s}_1)$ in the tangent space~\cite{Lee2002} of the switching boundary. From the differentiability with respect to initial conditions, the following relation holds for all $t$ outside the $\epsilon$ neighborhoods of the moments of switching, 
\begin{align}
		{\bm X}_2(t) = {\bm X}_1(t) + \epsilon {\bf \Xi}_{{\bm s}_1}(t) {\bm h} + O(\epsilon^2), \label{eq. errbyinit}
\end{align}
where ${\bm X}_1(t)$ and ${\bm X}_2(t)$ are the continuous part of the solutions with the initial conditions ${\bm s}_1$ and ${\bm s}_2$, respectively, and ${\bf \Xi}_{{\bm s}_1}(t) \in \mathbb{R}^{N\times N}$ denotes the initial condition sensitivity matrix with the initial state ${\bm s}_1$. Since ${\bm s}_1$ and ${\bm s}_2$ are taken from $\mathcal{A}$, which is a subset of the basin of attraction $U$ of the hybrid limit cycle $\chi$, the solutions ${\bm X}_1, {\bm X}_2$ relax to $\chi$, which we denote as ${\bm X}_{0,1}, {\bm X}_{0,2}$ in this section. One can easily see that the following relation holds: 
	\begin{align}
		{\bm X}_{0,2}(t) \! =\! {\bm X}_{0,1}(t)\! + \! \int_0^{\theta({\bm s}_2) - \theta({\bm s}_1)}{\! \! \! \! {\bm F}(k,{\bm X}_{0,1}(t \! + \! t'))dt'}. \label{eq. errbyint}
	\end{align}
	Since ${\bm X}_{0,1}(t)$ is not an equilibrium, we can assume that the first entry $F_1(k,{\bm X}_{0,1}(t))$ of ${\bm F}(k,{\bm X}_{0,1}(t))$ is nonzero without loss of generality and denote its absolute value as $|F_1(k,{\bm X}_{0,1}(t))| = d$. 
	Considering the continuity of the continuous part of the solution with respect to the initial condition and the continuity of the vector field ${\bm F}(k,{\bm X})$ with respect to the continuous variable ${\bm X}$, there exists $\epsilon' >0$ such that for any $\epsilon < \epsilon'$, the inequality $|F_1(k,{\bm X}_{0,1}(t + t'))| \ge d/2$ holds for $t' \in [0, \theta({\bm s}_2) - \theta({\bm s}_1)]$.
Hence we can obtain the following inequality for $\epsilon < \epsilon'$: 
\begin{align}
	&|{\bm X}_{0,2}(t)-{\bm X}_{0,1}(t)| = \left| \int_0^{\theta({\bm s}_2) - \theta({\bm s}_1)}{\! \! \! \! {\bm F}(k,{\bm X}_{0,1}(t + t'))dt'}\right| \cr
	&= |{\bm d}||\theta({\bm s}_2) - \theta({\bm s}_1)| 
	\ge \frac{d}{2}|\theta({\bm s}_2) - \theta({\bm s}_1)| \label{eq. errbyintrestated}
\end{align}
where ${\bm d}$ is a constant vector obtained by applying the mean value theorem to each entry of ${\bm F}(k,{\bm X})$. 
From Eqs.~(\ref{eq. errbyinit}) and (\ref{eq. errbyintrestated}), we can show the continuity of $\theta$ as follows: 
\begin{align}
	\lim_{{\bm s}_2 \to {\bm s}_1} \ {\! \! \!  |\theta({\bm s}_2)\!  - \! \theta({\bm s}_1)|}\! \le \! \lim_{\epsilon \to 0}{\! \left(\left|\frac{2\epsilon}{d}{\bf \Xi}_{{\bm s}_1}(t){\bm h}\right|\! + \! O(\epsilon^2)\! \! \right)}\! = \!  0.
\end{align}
The continuity assures that $\theta({\bm s}_2) - \theta({\bm s}_1)$ is $O(\epsilon)$. Therefore, we can restate Eq.~(\ref{eq. errbyint}) as
\begin{align}
	&{\bm X}_{0,2}(t) 
	= {\bm X}_{0,1}(t) + (\theta({\bm s}_2) - \theta({\bm s}_1)){\bm F}(k,{\bm X}_{0,1}(t)) + O(\epsilon^2). \label{eq. errbyinitrestated}
\end{align}
From Eqs.~(\ref{eq. errbyinit}) and (\ref{eq. errbyinitrestated}), we can obtain
\begin{align}
	\theta({\bm s}_2) \! - \! \theta({\bm s}_1) \! = \! \frac{\epsilon {\bm F}^{\mathrm{\dag}}(k,{\bm X}_{0,1}(t)){\bf \Xi}_{{\bm s}_1}(t){\bm h}}{{\bm F}^{\mathrm{\dag}}(k,{\bm X}_{0,1}(t)){\bm F}(k,{\bm X}_{0,1}(t))} \! + \! O(\epsilon^2). \label{eq. errphase}
\end{align}

Now we consider the asymptotic property of ${\bf \Xi}_{{\bm s}_1}(t)$. 
The assumption (C2) in Appendix~\ref{sec. assumptions} assures that for sufficiently small $\epsilon$, there exists $t_*$ such that the discrete states of ${\bm s}_1(t_*+nT)$ and ${\bm s}_2(t_*+nT)$ are the same and invariant for all $n \in \mathbb{Z}_{\ge 0}$ for any ${\bm h}$. 
We define a time-$T$ map ${\bm P}: \mathbb{R}^N \to \mathbb{R}^N$ as
\begin{align}
	{\bm P}({\bm X}(t)) = {\bm X}(t+T),
\end{align}
 and its $n$-fold composition as ${\bm P}^n$. 
 By a similar argument to the proof of the Lemma in Appendix A in~\cite{guckenheimer75}, we can show that the sequences $\{ {\bm P}^n({\bm X}_1(t_*))\}$ and $\{ D{\bm P}^n({\bm X}_1(t_*))\}$ are convergent and that $\lim_{n\to \infty}{{\bm P}^n({\bm X}_1(t_*))} = {\bm X}_0(\theta_*)$ and $\lim_{n\to \infty}{(D{\bm P})\circ( {\bm P}^{n-1}({\bm X}_1(t_*))} = {\bf M}(\theta_*)$. Here, $\theta_*$ is a unique phase value that depends on ${\bm X}_1(t_*)$, and the monodromy matrix ${\bf M}(\theta_*)$ is defined in Appendix~\ref{sec. linearstability}. 

Let us define ${\bm Q}(t_*) \equiv \lim_{n\to \infty}{D{\bm P}^n({\bm X}_1(t_*))} $. One can easily see that ${\bf M}(\theta_*){\bm Q}(t_*) = {\bm Q}(t_*)$. 
Using this relation and the fact shown in~\cite{khan2011sensitivity}, 
\begin{align}
	\lim_{n\to \infty}{{\bf M}^n(\theta_*)} = {\bm F}(I_0(\theta_*),{\bm X}_0(\theta_*))\otimes {\bm v}(\theta_*),
\end{align}
where ${\bm v}(\theta_*) \in \mathbb{R}^N$ is a left eigenvector of ${\bf M}(\theta_*)$ corresponding to the eigenvalue unity, one can obtain
\begin{align}
	&\lim_{n\to \infty}{{\bf \Xi}_{{\bm s}_1}(t_*+nT)}=	{\bm Q}(t_*){\bf \Xi}_{{\bm s}_1}(t_*) \cr
	&=\lim_{n\to \infty}{{\bf M}^n(\theta_*)}{\bm Q}(t_*){\bf \Xi}_{{\bm s}_1}(t_*) \cr
	&= {\bm F}(I_0(\theta_*),{\bm X}_0(\theta_*))\otimes {\bm v}(\theta_*){\bm Q}(t_*){\bf \Xi}_{{\bm s}_1}(t_*).
\end{align}
Note that ${\bm v}(\theta_*)$ also satisfies the condition ${\bm v}(\theta_*) \cdot {\bm F}(I_0(\theta_*),{\bm X}_0(\theta_*)) = 1$, and, hence, in fact, it is the phase sensitivity function evaluated at $\theta = \theta_*$ as shown in Appendix~\ref{sec. derivation_adj}.

Using ${\bm v}(\theta_*)$, ${\bm Q}(\theta_*)$ and ${\bf \Xi}_{{\bm s}_1}(\theta_*)$, we can rewrite Eq.~(\ref{eq. errphase}) as
\begin{align}
	\theta({\bm s}_2) - \theta({\bm s}_1) &= \epsilon {\bm v}^{\dag}(\theta_*)	{\bm Q}(t_*){\bf \Xi}_{{\bm s}_1}(t_*){\bm h} + O(\epsilon^2) 
	= \epsilon {\bm v}^{\dag}(\theta_*){\bm R}(\theta_*;{\bm s}_1){\bm h} + O(\epsilon^2),
\end{align}
where ${\bm R}(\theta_* ;{\bm s}_1) \equiv {\bm Q}(t_*){\bf \Xi}_{{\bm s}_1}(t_*)$ is introduced to emphasize that it depends only on $\theta_*$ and ${\bm s}_1$. 
As shown below, ${\bm v}^{\dag}(\theta_*){\bm R}(\theta_*;{\bm s}_1)$ on the right-hand side does not depend on the choice of $\theta_*$. 
One can see that ${\bm R}(\theta + \theta';{\bm s}_1) = {\bm \Xi}(\theta';\theta){\bm R}(\theta;{\bm s}_1)$ and that ${\bm v}(\theta) = {\bm \Psi}(-\theta';\theta+\theta'){\bm v}(\theta+\theta')$, where ${\bm \Xi}(\cdot ; \theta)$ and ${\bm \Psi}(\cdot; \theta)$ are defined in Appendix~\ref{sec. linearstability} and~\ref{sec. derivation_adj}, respectively. 
The latter equality follows from the fact shown in Appendix~\ref{sec. derivation_adj} that ${\bm v}(\theta)$ is a periodic solution of the adjoint linear system Eq.~(\ref{eq. psiadjvec}, \ref{eq. psiadjvecjump}). 
Similarly to Eq.~(\ref{eq. monoadj}), we can show ${\bm \Xi}(\theta';\theta) = {\bm \Psi}^{\dag}(-\theta';\theta+\theta')$. 
Thus,
\begin{align}
	&{\bm v}^{\dag}(\theta \! + \! \theta'){\bm R}(\theta \! + \! \theta';{\bm s}_1) \!= \! {\bm v}^{\dag}(\theta \! + \! \theta'){\bm \Xi}(\theta';\theta){\bm R}(\theta;{\bm s}_1) \cr
	&= \left( {\bm \Psi}(-\theta';\theta+\theta'){\bm v}(\theta+\theta') \right)^{\dag} {\bm R}(\theta;{\bm s}_1) \cr
	& = {\bm v}^{\dag}(\theta){\bm R}(\theta;{\bm s}_1), 
\end{align}
and we finally obtain
\begin{align}
	&\lim_{{\bm s}_2 \to {\bm s}_1}{\frac{|\theta({\bm s}_2) - \theta({\bm s}_1) - {\bm S}^{\dag}({\bm s}_1)({\bm X}_2(0) - {\bm X}_1(0))|}{|{\bm X}_2(0) - {\bm X}_1(0)|}} 
	= \lim_{\epsilon \to 0}{ O(\epsilon)} = 0, 
\end{align}
where we defined ${\bm S}^{\dag}({\bm s}_1) \equiv {\bm v}^{\dag}(\theta_*){\bm R}(\theta_*;{\bm s}_1)$.
This concludes the proof of the total differentiability (and one-sided differentiability at switching boundaries) of the phase.

The definition of the phase guarantees that the relation (Eq.~$(7)$ in the main article) 
\begin{align}
	\dot{\theta}(t) &= \dot{\Theta}(I(t), {\bm X}(t)) 
	= \nabla \Theta (I(t), {\bm X}(t)) \cdot {\bm F}(I(t),{\bm X}(t)) = 1
	\label{eq. phasedef app}
\end{align}
holds for an unperturbed oscillator for almost all $t$ (excluding the set of the moments of switching, which has zero Lebesgue measure) and for ${\bm X(t)}\in \tilde{U}$.
From this relation, $\nabla \Theta$ is nonzero everywhere on in $\tilde{U}$.
Therefore, by using the implicit function theorem, we can show that each connected component of the subset of the conditional isochron $\tilde{W}_k(\theta) = \{{\bm X} \: | \: {\bm X} \in W_k(\theta) \cap (k,{\bm X}) \in \tilde{U} \}$ is an $(N-1)$-dimensional smooth submanifold embedded in $\mathbb{R}^{N}$.

\section{Approximation of the phase dynamics
\label{sec. approxphase}}

Using the chain rule, the phase dynamics of the weakly perturbed hybrid oscillator described by Eq.~($8$) in the main article is given by
\begin{align}
	\dot{\theta}(t) &= \dot{\Theta}(I(t), {\bm X}(t)) 
	= 1  + \epsilon \nabla \Theta (I(t), {\bm X}(t)) \cdot {\bm p}(I(t), {\bm X}(t), t).
\end{align}
This is still not a closed equation in the phase $\theta$ because the map $\Theta:U\to \mathbb{T}^1$ is not injective. To obtain a closed equation, we assume the magnitude $\epsilon$ of the perturbation to be sufficiently small and approximate $\nabla \Theta (I(t), {\bm X}(t))$ and ${\bm p}(I(t), {\bm X}(t), t)$ by replacing $I(t)$ with $I_{0}(\theta(t))$ and ${\bm X}(t)$ with ${\bm X}_{0}(\theta(t))$~\cite{kuramoto2003chemical}.
We can then obtain the following approximate phase equation closed in $\theta$ at the lowest order:
\begin{align}
	\dot{\theta}(t) &= \dot{\Theta}(I(t), {\bm X}(t)) \cr
&= 1  \! + \! \epsilon \nabla \Theta (I_{0}(\theta), {\bm X}_0(\theta)) \cdot  {\bm p}(I_{0}(\theta), {\bm X}_0(\theta), t)\! + \! O (\epsilon^2 ) \cr
&= 1 + \epsilon {\bm Z}(\theta) \cdot {\bm p}(I_{0}(\theta), {\bm X}_0(\theta), t) + O (\epsilon^2 ),
\label{eq. axreducedeq}
\end{align}
where we defined the phase sensitivity function ${\bm Z}(\theta) = \nabla \Theta (I_0(\theta), {\bm X}_0(\theta))$ characterizing the linear response property of the oscillator phase to weak external perturbations.

We interpret Eq.~(\ref{eq. axreducedeq}) as a suitably regularized, multivalued system, such as the Filippov system~\cite{Filippov2013,Cortes2012discontinuous}, since some important solutions can not be obtained in the classical Carath\'eodory sense~\cite{Filippov2013}. For example, a stable stationary solution at the point of discontinuity of the right-hand side of Eq.~(\ref{eq. axreducedeq}) is, in general, not a Carath\'eodory solution, because it is required to satisfy Eq.~(\ref{eq. axreducedeq}) for almost all $t$ by definition, but the desired stationary solution may not satisfy Eq.~(\ref{eq. axreducedeq}) for all $t$.  

When the perturbation ${\bm p}$ is locally bounded, we can introduce the Filippov solution to Eq.~(\ref{eq. axreducedeq}) as an absolutely continuous map $\theta(t): \mathbb{R} \to \mathbb{T}^1$, which satisfies the following differential inclusion:
\begin{align}
	\dot{\theta}(t) \in 1 + \epsilon{\bm G}(\theta, t)
	\label{eq. phase_di}
\end{align}
for almost all $t$, where ${\bm G}(\theta, t)$ is a set of closed convex combinations of ${\bm Z}(\theta - 0)\cdot {\bm p}(I_0(\theta -0),{\bm X}_0(\theta - 0),t)$ and ${\bm Z}(\theta + 0)\cdot {\bm p}(I_0(\theta +0),{\bm X}_0(\theta +0),t)$. 
Obviously, the Filippov system Eq.~(\ref{eq. phase_di}) allows stationary solutions at the point of discontinuity described above. 
Note that the above Filippov regularization of the system can also produce physically meaningless solutions. For example, the Filippov system admits an evidently unfeasible, unstable stationary solution staying at the point of discontinuity of the right-hand side of Eq.~(\ref{eq. axreducedeq}).
This kind of solution, called a parasite solution~\cite{Krbec1986}, should be carefully omitted. 
See~\cite{Filippov2013,Cortes2012discontinuous} for sufficient conditions for the existence and uniqueness of solution to the Filippov system. 

When the perturbation is not locally bounded, for instance, when it includes the Dirac $\delta$ function, we need to consider a physically relevant solution, which is generally not absolutely continuous, by employing suitable formulations such as impulse differential inclusions~\cite{aubin2002impulse} or measure driven differential inclusions~\cite{silva1996measure}.
Though we do not consider such a special situation in this study,
if an impulsive input is applied at the moment of switching of the discrete states, it requires a special attention because the choice of the value of the integrand at the atom of the driving measure crucially affects the solution.  

Consider the case where an impulsive input ${\bm p}(\cdot,\cdot,t) = {\bm c}\delta(t-\tau)$, where $\delta(\cdot)$ is Dirac's delta function, is applied at the moment of switching $t=\tau$. 
If the impulsive input is applied when the state $(I_0(\theta(\tau)),{\bm X}_0(\theta(\tau)))$ is on the switching plane,
there are two possible cases:
(a) $(\nabla L_{I_0(\theta(\tau))}({\bm X}_0(\theta(\tau)))\cdot{\bm c})(\nabla L_{I_0(\theta(\tau))}({\bm X}_0(\theta(\tau)))\cdot{\dot{\bm X}}_0(\theta(\tau)))> 0$, i.e., the perturbation and the vector field point in the same direction in the tangent space~\cite{Lee2002} of the switching boundary or (b) otherwise. 
In the case (a), the one-sided derivative of the phase function ${\bf \Theta}$ in the direction of ${\bm c}$ is undefined.
If we redefine the hybrid automaton Eq.~(1)-(3) in the main article, for example, by replacing the switching surface with the switching region ${\bm \Pi}_{ij} = \{{\bm X} \ | \ L((i,j),{\bm X}) \le 0 \}$, we can introduce a one-sided derivative of ${\bf \Theta}$ in the direction of ${\bm c}$. However, in general, it does not coincide with the phase sensitivity function obtained from the proposed adjoint method. Hence, this situation requires special treatments beyond the scope of this study. 
In the case (b), we can adopt ${\bm Z}(\theta(\tau) - 0)$ as the phase sensitivity function.
When the input is added immediately after the reset of the discrete state, which can also be interpreted as a perturbation to the transition function as ${\bm \Phi}_{I_0(\theta(\tau))}({\bm X}_0(\theta(\tau)))+{\bm c}$, we can adopt ${\bm Z}(\theta(\tau) + 0)$ as the phase sensitivity function. 

\section{Adjoint equation for the phase sensitivity function \label{sec. derivation_adj}}

An adjoint linear system to Eqs. (\ref{eq. vareqsmooth}) and (\ref{eq. varjump}) can also be introduced as
\begin{align}
	&\dot{{\bf \Psi}}(t) = - {\bf A}^{\dag}(k,t) \ {\bf \Psi}(t) 
	\mbox{\ \ for\ \ } t \ (\mathrm{mod}  \ T) \in (\tau_{k - 1}({\bm s}^*),\tau_{k}({\bm s}^*)), \label{eq. adjode} \\
		&{\bf \Psi}(t) = {\bf C}_{k}^{\dag} \ {\bf \Psi}(t+0)
	\mbox{\ \ at\ \ } t \ (\mathrm{mod} \ T) = \tau_{k}({\bm s}^*), \label{eq. adjjump}
\end{align}
with the initial condition
\begin{align}
	{\bf \Psi}(0) &= {\bf I}. \label{eq. adjinit}
\end{align}
Note that the above adjoint system can be integrated only backward in time because ${\bf C}_k$ can be singular.
The state transition matrix ${\bf H}_k(t,s)$ of the variational equation (see Eqs.~(\ref{eq. transfer}) and (\ref{eq. odetransfer}) for the definition) can always be inverted when $t, s \in [\tau_{k - 1}({\bm s}^*) + nT,\tau_{k}({\bm s}^*) + nT]$ for each $k = 1, ..., m_0$. 
In each time domain, we can obtain 
\begin{align}
	&(\dot{\bf H}_k^{-1})^{\dag}(t,s) = -{\bf A}^{\dag}(k,t) \ ({\bf H}_k^{-1})^{\dag}(t,s)
	\mbox{\ \ for\ \ }	t, s \in [\tau_{k -1}({\bm s}^*)+nT,\tau_{k}({\bm s}^*)+nT] \label{eq. adjtransfer}
\end{align}
by differentiating the identity
\begin{align}
{\bf H}^{-1}_k(t,s)\cdot {\bf H}_k(t,s) = {\bf I}
\end{align}
with $t$.
Using the periodicity ${\bf H}_k(t,s)={\bf H}_k(t+nT,s+nT)$, we can formally consider that Eq.~(\ref{eq. adjtransfer}) holds in the negative time domain. 
Thus, the matrix $({\bf H}_{k}^{-1})^{\dag}(t,s)$ satisfies the adjoint system (\ref{eq. adjode}) within each time interval 
with the initial condition $({\bf H}_{k}^{-1})^{\dag}(s,s) = {\bf I}$. 
Thus, $({\bf H}_{k}^{-1})^{\dag}(t,s)$ is a state transition matrix of the adjoint Eqs.~(\ref{eq. adjode}) satisfying
\begin{align}
					&{\bf \Psi}(t) = ({\bf H}_{k}^{-1})^{\dag}(t,s)	{\bf \Psi}(s) \cr
			&\mbox{\ \ for\ }  t, s \! \in \! (\tau_{k-1}({\bm s}^*)\! - \!(n\! +\! 1)T,\tau_{k}({\bm s}^*)\! -\! (n\! +\! 1)T], k\! \in \! \mathcal{M}_0  \cr		
				&\mbox{\ \ and for\ \ } t, s \in (\tau_{m_0}({\bm s}^*)-(n+1)T,-nT]
				\end{align}
We define a monodromy matrix of the adjoint system as ${\bf M}_\mathrm{adj}={\bf \Psi}(-T)$. It is easy to see that ${\bf M}_\mathrm{adj}$ can be expressed as 
\begin{widetext}
\begin{align}
{\bf M}_\mathrm{adj}  =  \left( \sideset{}{^+}\prod_{k=1}^{m_0}{({\bf H}_{m_0 + 1 - k}^{-1})^{\mathrm{\dag}}(\tau_{m_0 - k} ({\bm s}^*),\tau_{m_0 + 1 -k}({\bm s}^*)){\bf C}_{m_0 + 1 -k}^{\mathrm{\dag}}} \right) ({\bf H}_1^{-1})^{\dag}(\tau_{m_0}({\bm s}^*),T) \nonumber \\
=  \left( \sideset{}{^+}\prod_{k=1}^{m_0}{{\bf H}_{m_0 + 1 - k}^{\mathrm{\dag}}(\tau_{m_0 + 1 -k}({\bm s}^*),\tau_{m_0 - k} ({\bm s}^*)){\bf C}_{m_0 + 1 -k}^{\mathrm{\dag}}} \right) {\bf H}_1^{\mathrm{\dag}}(T,\tau_{m_0}({\bm s}^*)) = {\bf M}^{\mathrm{\dag}}. \label{eq. monoadj}
\end{align}
\end{widetext}
In the second equality, we used the relation ${\bf H}_k(t,s) = {\bf H}_k^{-1}(s,t)$.
Similarly, we consider adjoint linear systems corresponding to the systems with initial variations Eq.~(\ref{eq. defvariation}), whose solutions are ${\bf \Psi}(\cdot \ ; \theta)$, which satisfies
\begin{align}
	&\dot{{\bf \Psi}}(t;\theta) = - {\bf A}^{\dag}(k,t+\theta) \ {\bf \Psi}(t;\theta) 
	\mbox{\ \ for\ \ } t + \theta \ (\mathrm{mod}  \ T) \in (\tau_{k - 1}({\bm s}^*),\tau_{k}({\bm s}^*)), \\
		&{\bf \Psi}(t) = {\bf C}_{k}^{\dag} \ {\bf \Psi}(t+0)
	\mbox{\ \ at\ \ } t + \theta \ (\mathrm{mod} \ T) = \tau_{k}({\bm s}^*), \\ 
	&{\bf \Psi}(0;\theta) = {\bf I}.
\end{align}
It can be easily shown that ${\bf M}_\mathrm{adj}(\theta) = {\bf M}^{\dag}(\theta)$, where ${\bf M}_\mathrm{adj}(\theta) \equiv {\bf \Psi}(-T;\theta)$.

Since we assume that the hybrid limit cycle $\chi$ is linearly stable, the monodromy matrix ${\bf M}(\theta)$ has a single eigenvalue $1$ and all other eigenvalues are strictly inside the unit circle on the complex plane.
We denote the eigenvalues as $\lambda_{i}$ $(i=1, 2, ..., N)$ and the corresponding right eigenvectors as ${\bm u}_{i}(\theta)$, where $\lambda_{1} = 1$ and $|\lambda_{i}| < 1$ $(i=2, ..., N)$.
For simplicity, we hereafter consider the case where all eigenvalues are real and semisimple. A similar argument holds for the case of complex conjugates and eigenvalues with the generalized eigenspaces.
It can be shown that ${\bm u}_{i}(\theta)$ corresponding to the eigenvalue $\lambda_{i}$ with $|\lambda_{i}| < 1$ ($i=2, ..., N$) is tangent to the conditional isochron $W_{I_0(\theta )}(\theta )$  
at ${\bm X}_0(\theta)$, and that
the right eigenvector corresponding to $\lambda_1 = 1$ is tangent to the limit cycle $\chi$ at ${\bm s}_0(\theta)$.

Let ${\bm \eta}(t)$ be a variation vector from the unperturbed limit-cycle orbit ${\bm X}_{0}(t+\theta)$.
If the initial value ${\bm \eta}(0)=\epsilon {\bm h}$, where $|\epsilon| \ll 1$, given to the initial state ${\bm X}_0(\theta)$ is parallel to the eigenvector ${\bm u}_i(\theta)$ corresponding to $|\lambda_i|<1 (i=2,\cdots ,N)$, i.e., ${\bm \eta}(0) \propto {\bm u}_i(\theta)$, the linear response to the initial perturbation ${\bm \eta}(0)$ eventually vanishes, i.e.,
\begin{align}
	\lim_{n\to \infty}{{\bm \eta}(nT)}  &= \lim_{n\to \infty}{\epsilon{\bf \Xi}(nT;\theta){\bm h}} \cr
	&= \lim_{n\to \infty}{\epsilon {\bf M}(\theta)^n {\bm h}} = \lim_{n\to \infty}{\epsilon \lambda_i^n {\bm h}} = {\bm 0},
\end{align}
as the system state revolves around $\chi$.
Hence, the state $\lim_{n \to \infty} [ {{\bm X}_0(nT+\theta)+{\bm \eta}(nT)} ]$ can be approximated by ${\bm X}_0(\theta) + O(\epsilon^2)$, and it shares the same phase with ${\bm X}_0(\theta) + {\bm \eta}(0)$.
Using the total differentiability of the phase function ${\bf \Theta}$ with respect to the continuous state, the directional derivative of the phase in the direction ${\bm h}$ is obtained as 
\begin{align}
&\nabla{\bf \Theta}(I_0(\theta),{\bm X}_0(\theta))\cdot {\bm h} \cr
&= \lim_{\epsilon \to 0}{\frac{{\bf \Theta}(I_0(\theta),{\bm X}_0(\theta)+{\bm \eta}(0)) - {\bf \Theta}(I_0(\theta),{\bm X}_0(\theta))}{\epsilon}} \cr
&=  \lim_{\epsilon \to 0}{\frac{{\bf \Theta}(I_0(\theta),{\bm X}_0(\theta)+O(\epsilon^2)) - {\bf \Theta}(I_0(\theta),{\bm X}_0(\theta))}{\epsilon}} \cr
&= \lim_{\epsilon \to 0}{O(\epsilon)} = 0.
\end{align}
Therefore, ${\bm u}_i(\theta) (i=2,\cdots, N)$ is tangent to $W_{I_0(\theta)}(\theta)$ at ${\bm X}_0(\theta)$.
In contrast, if ${\bm \eta}(0)$ is parallel to ${\bm u}_1$, ${\bm \eta}(t)$ does not decay since ${\bf M}(\theta)^n{\bm \eta}(0)={\bm \eta}(0)$. 
In Ref.~\cite[Th. 4.2.]{khan2011sensitivity}, it is shown that ${\bm F}(I_0(\theta),{\bm X}_0(\theta))$ is a right eigenvector of ${\bf M}(\theta)$ associated with the eigenvalue of unity. Hence ${\bm u}_1(\theta)$ is parallel to ${\bm F}(I_0(\theta),{\bm X}_0(\theta))$, which means that it is tangent to the limit cycle $\chi$ at ${\bm s}_0(\theta)$. 

We denote the left eigenvector corresponding to eigenvalue $\lambda_{1} = 1$ of ${\bf M}(\theta)$ as ${\bm Z}(\theta)$.
Then, ${\bm Z}(\theta)$ is orthogonal to all right eigenvectors ${\bm u}_{i}(\theta)$ $(i=2, ..., N)$ with eigenvalues $|\lambda_{i}| < 1$, namely, ${\bm Z}(\theta)$ is normal to the submanifold $W_{I_0(\theta)}(\theta)$ at ${\bm s}_0(\theta)$ and is parallel to the gradient vector $\nabla {\bf \Theta}|_{{\bm s}_0(\theta)}$.
Thus, when ${\bm Z}(\theta)$ is normalized so that ${\bm Z}(\theta) = \nabla {\bf \Theta} |_{{\bm s}_0(\theta)}$ holds, it gives the linear response of the phase variable to an applied perturbation, hence we call it a phase sensitivity function.

In the following, we describe why one can obtain the phase sensitivity function from the adjoint equation. Since the system given by~(\ref{eq. adjode}) and (\ref{eq. adjjump}) is linear, the solution of the adjoint linear system
	\begin{align}
		&\dot{\bm \psi}(t) = -{\bf A}^{\dag}(k,t){\bm \psi}(t) 
		\mbox{\ \ for\ \ } t \ (\mathrm{mod}  \ T) \in (\tau_{k - 1}({\bm s}^*),\tau_{k}({\bm s}^*)), \label{eq. psiadjvec} \\
		&{\bm \psi}(t) = {\bf C}_{k}^{\dag} {\bm \psi}(t+0)  \mbox{\ \ at\ \ } t \ (\mathrm{mod} \ T) = \tau_{k}({\bm s}^*), \label{eq. psiadjvecjump}
	\end{align}
	where ${\bm \psi}(t) \in \mathbb{R}^N$, is given as ${\bm \psi}(t) = {\bf \Psi}(t){\bm \psi}(0)$.
If we write ${\bm \psi}(0)$ as ${\bm Z}_1 + {\bm Z}_2$, where ${\bm Z}_1$ is the projection of ${\bm \psi}(0)$ onto the space spanned by ${\bm Z}(0)$ and ${\bm Z}_2$ is the remainder, the backward-in-time asymptotic solution is
\begin{align}
	\lim_{n\to \infty}{{\bm \psi}(-nT)} &= \lim_{n\to \infty}{{\bf M}_\mathrm{adj}(0)^n({\bm Z}_1 + {\bm Z}_2)} 
	= \lim_{n\to \infty}{{\bf M}^{\dag}(0)^n({\bm Z}_1 + {\bm Z}_2)} = {\bm Z}_1, \label{eq. asymptadj}
\end{align}
because ${\bm Z}(0)$ is a right eigenvector of ${\bf M}^{\dag}(0)$ corresponding to the eigenvalue $\lambda_1 = 1$.
From Eq.~(\ref{eq. asymptadj}), one can see that the backward-in-time asymptotic solution is periodic, hence we write it as ${\bm \psi}_0(\theta)$ with the initial value ${\bm \psi}_0(0) = {\bm Z}_1$. 
The vector ${\bm \psi}_0(\theta)$ is parallel to ${\bm Z}(\theta)$ because the equalities ${\bf M}_\mathrm{adj}(\theta){\bm \psi}_0(\theta) = {\bf M}^{\dag}(\theta){\bm \psi}_0(\theta)  ={\bm \psi}_0(\theta)$ hold, where the last equality means that ${\bm \psi}_0(\theta)$ is a left eigenvector of ${\bf M}(\theta)$ with a corresponding eigenvalue of unity.
We normalize ${\bm \psi}(0)$ as follows: 
\begin{align}
	{\bm \psi}(0)\cdot {\bm F}(I_0(0),{\bm X}_0(0)) = 1. \label{eq. psinorm}
\end{align}
Since ${\bm F}(I_0(0),{\bm X}_0(0))$ is a right eigenvector of ${\bf M}(0)$ corresponding to the eigenvalue $\lambda_1 = 1$, ${\bm Z}_2 \cdot {\bm F}(I_0(0),{\bm X}_0(0)) =0$ holds, and we obtain ${\bm \psi}_0(0)\cdot {\bm F}(I_0(0),{\bm X}_0(0)) = {\bm Z}_1 \cdot {\bm F}(I_0(0),{\bm X}_0(0)) = 1$. Therefore, under the condition~(\ref{eq. psinorm}), ${\bm \psi}_0(0)$ is equal to the phase sensitivity function ${\bm Z}(0)$ at $\theta = 0$ because it satisfies the relation~(\ref{eq. phasedef}).

It can be shown that the normalization condition is satisfied for all $\theta$, i.e, ${\bm \psi}_0(\theta)\cdot {\bm F}(I_0(\theta),{\bm X}_0(\theta)) = 1$, if ${\bm \psi}(0)$ satisfies the above normalization condition at $t=0$, as follows.
Hereafter, we formally define ${\bm F}(I_0(t-nT),{\bm X}_0(t-nT))={\bm F}(I_0(t),{\bm X}_0(t))$, because $(I_0(t),{\bm X}_0(t))$ is a periodic solution. 
By differentiating $d{\bm X}_{0}(t)/dt = {\bm F}(I_{0}(t), {\bm X}_{0}(t))$ by $t$ within the smooth interval, we obtain
\begin{align}
	&\frac{d}{dt} \left( \frac{d{\bm X}_{0}(t)}{dt}  \right) \cr 
	&= \frac{d}{dt}{\bm F}(I_{0}(t), {\bm X}_0(t)) 
	={\bm A}(I_0(t), t) \left( \frac{d {\bm X}_{0}(t)}{dt} \right) \cr
	&=
	{\bm A}(I_{0}(t), t) {\bm F}(I_{0}(t), {\bm X}_0(t)).
\end{align}
Thus, ${\bm \xi}(t) = {\bm F}(I_{0}(t), {\bm X}_0(t))$ is a solution to the vector-valued version of the linearized system (\ref{eq. vareqsmooth},\ref{eq. varjump}),
\begin{align}
	\frac{d}{dt}{\bm F}(I_{0}(t), {\bm X}_0(t))\!=\!{\bm A}(I_0(t), t) {\bm F}(I_{0}(t), {\bm X}_0(t)),
\end{align}
from which we can derive 
\begin{align}
&\frac{d}{dt}({\bm \psi}(t) \cdot {\bm F}(I_{0}(t), {\bm X}_0(t))) \cr
&=
\frac{d}{dt}{\bm \psi}(t) \cdot {\bm F}(I_{0}(t), {\bm X}_{0}(t)) + {\bm \psi}(t) \cdot \frac{d}{dt} {\bm F}(I_{0}(t), {\bm X}_{0}(t)) \cr
&=
- {\bf A}(I_{0}(t), t)^{\dag} {\bm \psi}(t) \cdot {\bm F}(I_{0}(t), {\bm X}_0(t))) 
+{\bm \psi}(t) \cdot {\bf A}(I_{0}(t), t) {\bm F}(I_{0}(t), {\bm X}_0(t)))
=
0.
\end{align}
At the moment of switching ($t \ (\mathrm{mod} \ T) = \tau_{k}({\bm s}^*)$), the variation ${\bm \xi}(t) = {\bm F}(I_{0}(t), {\bm X}_0(t))$ changes as
\begin{align}
	{\bm F}(k+1, {\bm X}_0(t)) = {\bf C}_k {\bm F}(k, {\bm X}_0(t)) .
\end{align}
Thus,
\begin{align}
	&{\bm \psi}(t+0) \cdot {\bm F}(k+1, {\bm X}_0(t)) \cr
	&=\! {\bm \psi}(t+0) \cdot {\bf C}_k {\bm F}(k ,{\bm X}_0(t)) 
	\!= \! {\bf C}_k^{\dag}{\bm \psi}(t+0) \cdot {\bm F}(k ,{\bm X}_0(t)) \cr
	&= {\bm \psi}(t) \cdot {\bm F}(k ,{\bm X}_0(t)).
\end{align}
Therefore, the quantity ${\bm \psi}(t)\cdot {\bm F}(I_{0}(t), {\bm X}_0(t))$ is invariant under the backward time evolution of the system given by~(\ref{eq. psiadjvec}) and (\ref{eq. psiadjvecjump}).

Summarizing, we can obtain ${\bm Z}(\theta)$ by integrating the adjoint system (\ref{eq. psiadjvec},\ref{eq. psiadjvecjump}) backward in time from a initial condition that satisfies the normalization condition~(\ref{eq. psinorm}) 
until a periodic solution is obtained. In conventional smooth systems, this procedure is called the {\it adjoint method}~\cite{ermentrout2010mathematical}. 
\section{Averaging approximation and analysis of the synchronization dynamics
\label{sec. approxphasedyn}}

By differentiating both sides of Eq.~(13) in the main article with respect to time and substituting the phase equation Eq.~(\ref{eq. axreducedeq}), we obtain a nonautonomous system
\begin{align}
	\dot{\psi} = \epsilon[\Delta + {\bm Z}((T/T_\mathrm{ext})t + \psi)\cdot {\bm p}(t)]. \label{eq. apdnonautom}
\end{align}
The averaging approximation for weakly perturbed oscillators~\cite{kuramoto2003chemical,hoppensteadt1997weakly,ermentrout2010mathematical} provides the following autonomous system:
\begin{align}
	\dot{\psi} = \epsilon[\Delta + \Gamma(\psi)] \equiv J(\psi), \label{eq. apdaveraged}
\end{align}
where 
\begin{align}
	\Gamma(\psi) = \frac{1}{T_\mathrm{ext}}\int_0^{T_\mathrm{ext}}{{\bm Z}((T/T_\mathrm{ext})t + \psi)\cdot {\bm p}(t)}dt. \label{eq. apdcplf}
\end{align}

We here summarize some useful theorems for the analysis of the synchronization dynamics. 
Bogolyubov's second theorem~\cite{hale69,Bogolyubov61asymptotic+Mitropolsky67averaging,guckenheimer83nonlinear} affirms that the existence of a hyperbolic fixed point $\psi^*$, i.e., $J(\psi^*)=0$ and $J'(\psi^*)\neq 0$, of the averaged system Eq.~(\ref{eq. apdaveraged}) assures its corresponding unique hyperbolic periodic solution of the original system Eq.~(\ref{eq. apdnonautom}) evolving in the neighborhood of $\psi^*$ whose radius tends to zero together with $\epsilon$. 
The Eckhaus/Sanchez-Palencia theorem~\cite{Eckhaus75new+Sanchez75methode+Sanders07averaging} says that if $\psi^*$ is a stable hyperbolic fixed point of the averaged system, the solution of the nonaveraged system starting from the basin of attraction is estimated up to $O(\epsilon)$ by the averaged one with the same initial condition, which is uniformly valid over a semi-infinite time interval.
Samoilenko and Stanzhitskii~\cite{samoilenko06on} have given similar results under a less restrictive condition, where the hyperbolicity assumption in the Eckhaus/Sanchez-Palencia theorem is replaced by the asymptotic stability. 
These theorems mean that in each basin of attraction, the precise behavior of the original system can be captured by the averaged system.
Moreover, if the averaged system undergoes a saddle-node bifurcation at $\Delta = \Delta_c$ and $|\epsilon|$ is sufficiently small, the original system (in fact, its Poincar\'e map) also undergoes a saddle-node bifurcation at $\tilde{\Delta}_c$ near $\Delta_c$~\cite{guckenheimer83nonlinear}. 
Finally, from Bogolyubov's first theorem~\cite{hale69,Bogolyubov61asymptotic+Mitropolsky67averaging,guckenheimer83nonlinear}, even when the averaged system has no asymptotically stable fixed points, the solution of the nonaveraged system is estimated up to $O(\epsilon)$ on a time scale of order $O(1/\epsilon)$. 

Using the above theorems, synchronization dynamics of the oscillator can be easily understood from the $T$-periodic function $\Gamma(\psi)$ as follows: 
if the condition $\Delta \in [-\max{\Gamma(\psi)},-\min{\Gamma(\psi)}]$ is satisfied, Eq.~(\ref{eq. apdaveraged}) has at least one fixed point $\psi^*$ that satisfies $\Delta + \Gamma (\psi^* ) = 0$. When it is asymptotically stable, the oscillator is locked to the external forcing and the stable phase difference between the oscillator and the forcing is approximated by $\psi^{*}$. 
If there are two or more stable fixed points, the oscillator can synchronize with the periodic forcing at multiple phase differences depending on the initial condition. Each basin of attraction and convergence rate toward the stable phase differences can be estimated.  
Appearances and disappearances of stable phase differences, depending on the parameter of the frequency mismatch $\Delta$, can also be predicted from $\Gamma(\psi)$. 
When there are no stable phase differences, the phase slipping behavior occurs. The mean period of the phase slipping can also be estimated from $\Gamma(\psi)$. 

When one considers Eq.~(\ref{eq. apdnonautom}) as a regularized multivalued system as explained in Appendix~\ref{sec. approxphase},
the integral in Eq.~(\ref{eq. apdcplf}) should be interpreted in a suitable sense such as Aumann's~\cite{aumann65integrals}, and the differential equation Eq.~(\ref{eq. apdaveraged}) should be replaced by a differential inclusion.
See~\cite{Perestyuk2013averaging+Perestyuk2011differential} for the theories including analogues of the theorems mentioned above on the averaging approximation in systems with jumps and multivalued righthand sides. 
Note that the jumps in the solution in~\cite{Perestyuk2013averaging+Perestyuk2011differential} are assumed to occur when the solution collides with a switching surface described by $t = \tau(\psi)$ in the extended phase space, which is suitable to the case where one considers the injection locking to a periodic impulsive signal. 

Finally, when the mutual entrainment of pulse-coupled oscillators is analyzed~\cite{veryfast}, which we do not consider in this study, averaging approaches of the above kind should be generalized to the autonomous case. 
The averaging theory for autonomous systems with jumps has so far been limited to a specific cases~\cite{burd96resonant+newman15resonance}, and establishment of a general theory is desirable.  

\section{Direct method for measuring the phase sensitivity function} \label{sec. directmethod}

In the direct method, the $\beta$th element $Z^{\beta}(\theta)$ of ${\bm Z}(\theta)$ is computed as follows: 
First, we kick the system state $(I_{0}(\theta), {\bm X}_0(\theta))$ on the limit cycle $\chi$ by applying a weak impulsive perturbation $(0,\epsilon {\bm e}^{\beta})$.
We then evolve the orbit $(I(t), {\bm X}(t))$ from the perturbed initial condition $(I_{0}(\theta), {\bm X}_0(\theta) + \epsilon {\bm e}^{\beta})$.
After a long time, the perturbed orbit $(I(t), {\bm X}(t))$ returns sufficiently close to the limit cycle $\chi$ and the phase $\theta(I(t), {\bm X}(t))$ can be measured.
Because the phase difference between two unperturbed systems is time invariant, $\theta(I(t), {\bm X}(t)) - t_r$, where $t_r = t \mbox{\ mod\ } T$, is equal to the initial phase difference $\theta(I_{0}(\theta), {\bm X}_0(\theta) + \epsilon {\bm e}^{\beta}) - \theta$.  Thus, for sufficiently small $\epsilon$, we can calculate $Z^{\beta}(\theta)$ according to: 
\begin{align}
	Z^{\beta}(\theta)
	&= {\bm Z}(\theta) \cdot {\bm e}^{\beta}
	\approx \frac{ [ \theta(I_{0}(\theta), {\bm X}_0(\theta)+ \epsilon {\bm e}^{\beta}) - \theta ]}{\epsilon} 
	= \frac{ [\theta(I(t), {\bm X}(t)) - t_r] }{ \epsilon }.
\end{align}
In the direct method, the perturbation needs to be sufficiently small, as strong perturbations induce nonlinearity in the phase response.  However, too weak perturbations result in tiny phase responses, which are difficult to measure accurately.  Thus, the direct method is vulnerable to incorrect estimation of the phase response.
Moreover, the direct method requires much longer computation times than those for the adjoint method.  To calculate ${\bm Z}(\theta)$ at $m$ points on the limit cycle in hybrid dynamical systems with $N$-dimensional continuous states, it is necessary to repeat the above long-time evolution $m\times N$ times if we use the direct method.  In contrast, we need only a single long-time evolution in the adjoint method. Therefore, the adjoint method has a significant advantage in computing ${\bm Z}(\theta)$.

\section{Derivation of the negative logarithmic scaling law \label{sec. neglog}}

From Eq.~(\ref{eq. apdaveraged}), the period of phase slipping is estimated as
\begin{align}
	T_{\mathrm{slip}} = \left| \int_0^T{\frac{d\psi}{\epsilon [\Delta + \Gamma(\psi)]}} \right |. \label{eq. apptslip}
\end{align}
We suppose $\Gamma (\psi)$ has a maximum $-\Delta_c$ (the argument of the same kind holds for the minimum) at $\psi  = \psi^*$, and suppose the semiderivatives $\Gamma' (\psi^*-0)=\beta_1,\Gamma' (\psi^*+0)=\beta_2$ are nonzero. When $\Delta$ is sufficiently close to the critical value $\Delta_c$, $T_{\mathrm{slip}}$ is evaluated as
\begin{widetext}
\begin{align}
	T_{\mathrm{slip}} &\simeq -\frac{1}{\epsilon}\left( \int_0^{\psi^*}{\frac{d\psi}{\Delta - \Delta_c + \beta_1(\psi - \psi^*)}} + \int_{\psi^*}^{T}{\frac{d\psi}{\Delta - \Delta_c + \beta_2(\psi - \psi^*)}}\right) \cr
					&\simeq -\frac{1}{\epsilon}\left( \int_{-\infty}^{\psi^*}{\frac{d\psi}{\Delta - \Delta_c + \beta_1(\psi - \psi^*)}} + \int_{\psi^*}^{\infty}{\frac{d\psi}{\Delta - \Delta_c + \beta_2(\psi - \psi^*)}}\right) 
					\cr
	    &= -\frac{1}{\epsilon}\left( \frac{1}{\beta_1} - \frac{1}{\beta_2} \right)\ln{|\Delta - \Delta_c|}.
\end{align}
\end{widetext}
Hence, $T_{\mathrm{slip}}$ increases as $-\ln{|\Delta - \Delta_c|}$ when $\Delta \to \Delta_c$.

\begin{thebibliography}{99}

	\bibitem{Aranson2006patterns}
		I. S. Aranson and L. S. Tsimring,
		Rev. Mod. Phys. {\bf 78}, 641 (2006).
			
	\bibitem{Wolf99geometry}
		K. B. Wolf and G. Kr\"{o}tzsch,
		Eur. J. Phys. {\bf 16}, 14 (1995).

	\bibitem{Aihara2010theory}
		K. Aihara and H. Suzuki,
		Philos. Trans. R. Soc. Lond. A {\bf 368}, 4893 (2010).
		
	\bibitem{Holmes2006dynamics}
		P. Holmes, R. J. Full, D. Koditschek and J. Guckenheimer,
		SIAM Rev. {\bf 48}, 207 (2006).

	\bibitem{Macdonald09lateral}
		J. H. G. Macdonald,
		Proc. R. Soc. London A {\bf 465}, 1055 (2009).


	\bibitem{Andronov1987theory}
		A. A. Andronov, A. A. Vitt and S. E. Khaikin,
		{\it Theory of Oscillators}
		(Dover, New York, 2011).

		
	\bibitem{helbing2003modelling+Hespanha2001hybrid+Susuki09hybrid}
		D. Helbing,
		New J. Phys. {\bf 5}, 1 (2003); 
		J. P. Hespansa, S. Bohacek, K. Obraczka and J. Lee,
		Hybrid model of TCP congestion control, in {\it Hybrid systems: Computation and Control}
		(Springer, Berlin, 2001);
		Y. Susuki, Y. Takatsuji and T. Hikihara,
		IEICE Trans. Fund. Electr. Commun. Comput. Sci. {\bf 92}, 871 (2009).

		
	\bibitem{Makarenkov2012dynamics}
		O. Makarenkov and J. S. W. Lamb,
		Physica D {\bf 241}, 1826 (2012).
	
	\bibitem{Bernardo08Piecewise}
		M. di Bernardo, C. J. Budd, A. R. Champneys and P. Kowalczyk,
		{\it Piecewise-Smooth Dynamical Systems: Theory and Applications} 
		(Springer, London, 2008).
	
	\bibitem{coombes2012nonsmooth}
		S. Coombes, R. Thul and K. C. A. Wedgwood,
		Physica D {\bf 241}, 2042 (2012).

		
	\bibitem{McGeer90+Zimmermann07}
		T. McGeer,
		Int. J. Robot. Res. {\bf 9}, 62 (1990);
		K. Zimmermann and I. Zeidis,
		J. Theoret. Appl. Mech. {\bf 45}, 179 (2007).

	\bibitem{Banerjee01}
		S. Banerjee and G. C. Verghese (eds.), 
		{\it Nonlinear Phenomena in Power Electronics: Bifurcations, Chaos, Control, and Applications}
		(Wiley-IEEE press, New York, 2001).
	
	\bibitem{jenkins2013self}
		A. Jenkins,
		Phys. Rep. {\bf 525}, 167 (2013).
	
	\bibitem{Ijspeert98central+Strogatz05}
		A. J. Ijspeert,
		Neural Networks {\bf 21}, 642 (2008);
		S. Strogatz, D. Abrams, A. McRobie, B. Eckhardt and E. Ott,
		Nature {\bf 438}, 43 (2005).

	\bibitem{Dunnmon11power+Moss10low}
		J. A. Dunnmon, S. C. Stanton, B. P. Mann and E. H. Dowell,
		J. Fluid. Struct. {\bf 27}, 1182 (2011);
		S. Moss, A. Barry, I. Powlesland, S. Galea and G. P. Carman,
		Appl. Phys. Lett. {\bf 97}, 234101 (2010).
		
	\bibitem{tanaka2009self}
		H-A. Tanaka, H. Nakao and K. Shinohara,
		IEICE Electron. Expr. {\bf 6}, 1562 (2009).


	
	\bibitem{winfree2001geometry}
		A. T. Winfree,
		{\it The Geometry of Biological Time}
		(Springer, New York, 2001).
		
	\bibitem{kuramoto2003chemical}
		Y. Kuramoto,
		{\it Chemical Oscillations, Waves and Turbulence}
		(Dover, New York, 2003).
		
		
	\bibitem{hoppensteadt1997weakly}
		F. C. Hoppensteadt and E. M. Izhikevich,
		{\it Weakly Connected Neural Networks}
		(Springer, New York, 1997).

	\bibitem{Dorfler14synchronization+Harada10optimal+Zlotnik13optimal+Zlotnik16phaseselective}
		F. D\"{o}rfler and F. Bullo,
		Automatica {\bf 50}, 1539 (2014);
		T. Harada, H-A. Tanaka, M. J. Hankins and I. Z. Kiss,
		Phys. Rev. Lett. {\bf 105}, 088301 (2010);
		A. Zlotnik, Y. Chen, I. Z. Kiss, H-A. Tanaka and J-S. Li,
		Phys. Rev. Lett. {\bf 111}, 024102 (2013);
		A. Zlotnik, R. Nagao, I. Z. Kiss and J-S. Li,
		Nat. Commun. {\bf 7}, 10788 (2016).

	\bibitem{hale69}
		J. K. Hale,
		{\it Ordinary Differential Equations}
		(Wiley-Interscience, New York, 1969).
		
	\bibitem{guckenheimer75}
		J. Guckenheimer and S. Cruz,
		J. Math. Biol. {\bf 1}, 259 (1975).

	\bibitem{malkin1949methods}
		I. G. Malkin,
		{\it The Methods of Lyapunov and Poincar\'{e} in the Theory of Non-Linear Oscillations}
		(Gostexizdat, Moscow, 1949).
	
	\bibitem{ermentrout2010mathematical}
		G. B. Ermentrout and D. H. Terman,
		{\it Mathematical Foundations of Neuroscience}
		(Springer, New York, 2010).

	\bibitem{Yoshimura08+Teramae09+Goldobin10}
		K. Yoshimura and K. Arai,
		Phys. Rev. Lett. {\bf 101}, 154101 (2008);
		J-N. Teramae, H. Nakao and G. B. Ermentrout,
		Phys. Rev. Lett. {\bf 102}, 194102 (2009);
		D. S. Goldobin, J-N. Teramae, H. Nakao and G. B. Ermentrout,
		Phys. Rev. Lett. {\bf 105}, 154101 (2010).
		
	\bibitem{Novicenko12+Kotani12}
		V. Novic\~enko and K. Pyragas,
		Physica D {\bf 241}, 1090 (2012);
		K. Kotani, I. Yamaguchi, Y. Ogawa, Y. Jimbo, H. Nakao and G. B. Ermentrout,
		Phys. Rev. Lett. {\bf 109}, 044101 (2012).
		
	\bibitem{kawamurai2008collective+kori2009collective}
		Y. Kawamura, H. Nakao, K. Arai, H. Kori and Y. Kuramoto,
		Phys. Rev. Lett. {\bf 101}, 024101 (2008);
		H. Kori, Y. Kawamura, H. Nakao, K. Arai and Y. Kuramoto,
		Phys. Rev. E {\bf 80}, 036207 (2009).

	\bibitem{kawamura2013+nakao2014}
		Y. Kawamura and H. Nakao,
		Chaos {\bf 23}, 043129 (2013);
		H. Nakao, T. Yanagita, and Y. Kawamura,
		Phys. Rev. X {\bf 4}, 021032 (2014).

	\bibitem{kurebayashi2013phase}
		W. Kurebayashi, S. Shirasaka and H. Nakao,
		Phys. Rev. Lett. {\bf 111}, 214101 (2013).
	
	\bibitem{shaw2012phase}
		K. Shaw, Y. Park, H. Chiel and P. Thomas,
		SIAM J. Appl. Dyn. Syst. {\bf 11}, 350 (2012).

	\bibitem{demir2010+kawamura2015+mauroy2012}
		A. Demir, C. Gu and J. Roychowdhury,
		in {\it Proceedings of the 2010 IEEE/ACM International Conference on Computer-Aided Design, San Jose} (IEEE, New Jersey 2010), p. 292; 
		Y. Kawamura and H. Nakao,
		Physica D {\bf 295--296}, 11 (2015);
		A. Mauroy and I. Mezi\'{c},
		Chaos {\bf 22}, 033112 (2012).
			
	
	\bibitem{Ichinose98+Rabinovich99+Mauroy13}
		N. Ichinose, K. Aihara and K. Judd,
		Int. J. Bifurcation Chaos {\bf 8}, 2375 (1998);
		A. Rabinovich and I. Rogachevskii,
		Chaos {\bf 9}, 880 (1999);
		A. Mauroy, I. Mezi\'{c} and J. Moehlis,
		Physica D {\bf 261}, 19 (2013).
	
	\bibitem{izhikevich2000phase}
		E. M. Izhikevich,
		SIAM J. Appl. Math. {\bf 60}, 1789 (2000).
	
	\bibitem{park2013infinitesimal}
		Y. Park,
		Infinitesimal phase response curves for piecewise smooth dynamical systems, Master's thesis, 
		Case Western Reserve University, Cleveland, 2013.
	
	\bibitem{khan2011sensitivity}
		K. A. Khan, V. P. Saxena and P. I. Barton,
		SIAM J. Sci. Comput. {\bf 33}, 1475 (2011).

	\bibitem{Akhmet2005on+akhmet2010principles}
		M. U. Akhmet,
		Nonlinear Analysis {\bf 60}, 311 (2005); 
		{\it Principles of Discontinuous Dynamical Systems}
		(Springer, New York, 2010).

	\bibitem{Filippov2013}
		A. F. Filippov,
		{\it Differential Equations with Discontinuous Righthand Sides: Control Systems}
		(Springer, Berlin, 2013).

	\bibitem{Cortes2012discontinuous}
		J. Cort\'es,
		Control Systems, IEEE {\bf 28}, 36 (2008).

	\bibitem{Perestyuk2013averaging+Perestyuk2011differential}
		S. Klymchuk, A. Plotnikov and N. Skripnik,
		Physica D {\bf 241}, 1932 (2012);
		N. A. Perestyuk and N. V. Skripnik,
		Ukr. Math. J. {\bf 65}, 140 (2013);
		N. A. Perestyuk, V. A. Plotnikov, A. M. Samoilenko and N. V. Skripnik,
		{\it Differential Equations with Impulse Effects: Multivalued Right-hand Sides With Discontinuities}
		(Walter De Gruyter, Berlin, 2011).

		
	\bibitem{guckenheimer83nonlinear}
		J. Guckenheimer and P. Holmes,
		{\it Nonlinear Oscillations, Dynamical Systems, and Bifurcation of Vector Fields}
		(Springer, New York, 1983).


		
	\bibitem{izhikevich2010dynamical}
		E. M. Izhikevich,
		{\it Dynamical Systems in Neuroscience: The Geometry of Excitability and Bursting}
		(The MIT Press, Cambridge, 2010).

	\bibitem{veryfast}
		Y. Kuramoto,
		Physica D {\bf 50}, 15 (1991);
		A. Mauroy, P. Sacr\'{e} and R. Sepulchre,
		in {\it Proceedings of the 51st IEEE Conference on Decision and Control, Maui} (IEEE, New Jersey, 2012), p.~7171. 
		
	\bibitem{veryfast2}
		D. Somers and N. Kopell,
		Biol. Cybern. {\bf 68}, 393 (1993);

	\bibitem{garcia1998simplest}
		M. Garcia, A. Chatterjee, A. Ruina and M. Coleman,
		J. Biomech. Eng. {\bf 120}, 281 (1998).


	\bibitem{goswami1996compass}
		A. Gosmami, B. Thuilot and B. Espiau,
		INRIA RR-2996, (1996).


	\bibitem{Schaft00}
		A. J. van der Schaft and H. Schumacher,
		{\it An Introduction to Hybrid Dynamical Systems}
		(Springer, Berlin, 2000).	
	
	\bibitem{jeffrey2011nondeterminism}
		M. R. Jeffrey,
		Phys. Rev. Lett. {\bf 106}, 254103 (2011).
	
	\bibitem{horn2012}
		R. A. Horn and C. R. Johnson,
		{\it Matrix Analysis}
		(Cambridge University Press, Cambridge, 2012).

	\bibitem{biemond2013tracking}
		J. J. B. Biemond, N. van de Wouw, W. P. M. H. Heemels and H. Nijmeijer,
		IEEE Trans. Autom. Control {\bf 58}, 876 (2013).
	

		
	\bibitem{broucke2002continuous}
		M. Broucke and A. Arapostathis,
		Syst. Control Lett. {\bf 47}, 149 (2002).
		
	\bibitem{goebel2006solutions}
		R. Goebel and A. R. Teel,
		Automatica {\bf 42}, 573 (2006).
	
	\bibitem{simic2005towards}
		S. N. Simic. K. H. Johansson, J. Lygeros and S. S. Sastry,
		Dyn. Disc. Cont. Impuls. Syst., Series B {\bf 12}, 649 (2005).


		
	\bibitem{burden2014hybrid}
		S. A. Burden,
		A hybrid dynamical systems theory for legged locomotion, Ph.~D thesis, 
		University of California, Berkeley, 2014. 

	\bibitem{Lee2002}
		J. M. Lee,
		{\it Introduction to Smooth Manifolds}
		(Springer, New York, 2002).


	\bibitem{Krbec1986}
		P. Krbec,
		On nonparasite solutions, in {\it Equadiff 6}
		(Springer, Berlin, 1986).
		
	\bibitem{aubin2002impulse}
		J.-P. Aubin, J. Lygeros, M. Quincampoix and S. Sastry,
		IEEE Trans. Autom. Control {\bf 47}, 2 (2002). 

		
	\bibitem{silva1996measure}
		G. N. Silva and R. B. Vinter,
		J. Math. Anal. Appl. {\bf 202}, 727 (1996).
	
		

	\bibitem{Bogolyubov61asymptotic+Mitropolsky67averaging}
		N. N. Bogolyubov and I. A. Mitropolsky,
		{\it Asymptotic Methods in the Theory of Nonlinear Oscillations}
		(Gordon and Breach, New York, 1945);
		I. A. Mitropolsky,
		Int. J. Non-Linear Mech. {\bf 2}, 69 (1967).
		
		
			
	\bibitem{Eckhaus75new+Sanchez75methode+Sanders07averaging}
		W. Eckhaus,
		J. Math. Anal. Appl. {\bf 49}, 575 (1975);
		E. Sanchez-Palencia,
		C. R. Acad. Sci. S{\'e}rie A-B {\bf 280}, 105 (1975);
		J. A. Sanders, F. Verhulst and J. Murdock,
		{\it Averaging Methods in Nonlinear Dynamical Systems}
		(Springer, New York, 2007).
		
	\bibitem{samoilenko06on}
		A. M. Samoilenko and A. N. Stanzhitskii,
		Differ. Equ. {\bf 42}, 505 (2006).
		
		
	\bibitem{aumann65integrals}
		R. J. Aumann,
		J. Math. Anal. Appl. {\bf 12}, 1 (1965).

	\bibitem{burd96resonant+newman15resonance}
		V. S. Burd,
		Int. J. Non-linear Mech. {\bf 32}, 1143 (1997);
		J. Newman and O. Makarenkov,
		Nonlinear Dyn. {\bf 79}, 111 (2015).

\end{thebibliography}

\end{document}